\documentclass[twocolumn]{aastex631} 

\defcitealias{Q22}{Q22}
\defcitealias{K22}{G22}
\defcitealias{kamKinematicsMassDistribution2017}{K17}
\defcitealias{smercinaM33STRUCTURE2023}{S23}

\shorttitle{TREX: High-Dispersion Stellar Kinematics}
\shortauthors{Cullinane et al.}

\begin{document}

\title{TREX: Kinematic Characterisation of a High-Dispersion Intermediate-Age Stellar Component in M33}

\correspondingauthor{L. R. Cullinane}
\email{lcullin4@jhu.edu}

\author[0000-0001-8536-0547]{L. R. Cullinane}
\affiliation{Department of Physics and Astronomy, Johns Hopkins University, Baltimore, MD 21218, USA}

\author[0000-0003-0394-8377]{Karoline M. Gilbert}
\affiliation{Department of Physics and Astronomy, Johns Hopkins University, Baltimore, MD 21218, USA}
\affiliation{Space Telescope Science Institute, 3700 San Martin Dr., Baltimore, MD 21218, USA}

\author[0000-0001-8867-4234]{Puragra Guhathakurta}
\affiliation{UCO/Lick Observatory, Department of Astronomy \& Astrophysics, University of California Santa Cruz, 1156 High Street, Santa Cruz,
California 95064, USA}

\author[0000-0001-8481-2660]{A. C. N. Quirk}
\affiliation{Department of Astronomy, Columbia University, New York, NY, USA}

\author[0000-0002-9933-9551]{Ivanna Escala}
\altaffiliation{Carnegie-Princeton Fellow}
\affiliation{Department of Astrophysical Sciences, Princeton University, 4 Ivy Lane, Princeton, NJ, 08544 USA}
\affiliation{The Observatories of the Carnegie Institution for Science, 813 Santa Barbara St, Pasadena, CA 91101}

\author[0000-0003-2599-7524]{Adam Smercina}
\affiliation{Department of Astronomy, University of Washington, Box 351580, Seattle, WA 98195-1580, USA}

\author[0000-0002-7502-0597]{Benjamin F. Williams}
\affiliation{Department of Astronomy, University of Washington, Box 351580, Seattle, WA 98195-1580, USA}

\author[0000-0002-9599-310X]{Erik Tollerud}
\affiliation{Space Telescope Science Institute, 3700 San Martin Dr., Baltimore, MD 21218, USA}

\author[0009-0001-5335-2814]{Jessamine Qu}
\affiliation{Department of Computer Science, Carnegie Mellon University, 5000 Forbes Ave, Pittsburgh, PA 15213, USA}

\author{Kaela McConnell}
\affiliation{Department of Astronomy, Yale University, New Haven, CT 06520, USA}

\begin{abstract}
The dwarf galaxy Triangulum (M33) presents an interesting testbed for studying stellar halo formation: it is sufficiently massive so as to have likely accreted smaller satellites, but also lies within the regime where feedback and other “in-situ” formation mechanisms are expected to play a role. In this work, we analyse the line-of-sight kinematics of stars across M33 from the TREX survey with a view to understanding the origin of its halo. We split our sample into two broad populations of varying age, comprising 2032 “old” red giant branch (RGB) stars, and 671 “intermediate-age” asymptotic giant branch (AGB) and carbon stars. We find decisive evidence for two distinct kinematic components in both old and intermediate-age populations: a low-dispersion ($\sim22$~km s$^{-1}$) disk-like component co-rotating with M33’s \ion{H}{1} gas, and a significantly higher-dispersion component ($\sim50-60$~km~s$^{-1}$) which does not rotate in the same plane as the gas and is thus interpreted as M33’s stellar halo. While kinematically similar, the fraction of stars associated with the halo component differs significantly between the two populations: this is consistently $\sim$10\% for the intermediate age population, but decreases from $\sim34$\% to $\sim$10\% as a function of radius for the old population. We additionally find evidence that the intermediate-age halo population is systematically offset from the systemic velocity of M33 by $\sim25$~km~s$^{-1}$, with a preferred central LOS velocity of $\sim-155$~km~s$^{-1}$. This is the first detection and characterisation of an intermediate-age halo in M33, and suggests in-situ formation mechanisms, as well as potentially tidal interactions, have helped shaped it.
\end{abstract}



\section{Introduction} \label{sec:intro}
In the $\Lambda$Cold Dark Matter ($\Lambda$CDM) paradigm, galaxies form through hierarchical accretion, with smaller satellites accreted and torn apart by larger hosts to produce extended, diffuse, and kinematically hot stellar halos \citep[e.g.][]{whiteGalaxyFormationHierarchical1991,bullockTracingGalaxyFormation2005}. Studying these stellar halos therefore provides key insight into the history and evolution of a galaxy over cosmic time. This is clearly evidenced in observations of massive galaxies like the Milky Way (MW); for example, the ancient accretion of Gaia-Sausage-Enceladus forms a significant fraction of the MW’s inner halo \citep{helmiMergerThatLed2018}. As $\Lambda$CDM predicts this hierarchical growth occurs on all scales, similar accreted stellar halos are expected to exist around less massive systems as well. 

However, hierarchical accretion is not the only process through which extended, kinematically hot components can form. Major \citep{zolotovDUALORIGINSTELLAR2009} and minor mergers \citep[e.g.][]{quinnHeatingGalacticDisks1993,purcellHeatedDiscStars2010} can also displace and dynamically heat existing disk stars (i.e.\ “in-situ” populations) to produce kinematically hot components, with simulations suggesting such stars contribute substantially to the “inner halo” of galaxies \citep[e.g.][]{cooperFormationSituStellar2015,pillepichBUILDINGLATETYPESPIRAL2015}. Disk stars can also be dynamically heated through repeated scattering interactions with local density perturbations including giant molecular clouds \citep[e.g.][]{spitzerPossibleInfluenceInterstellar1951}, spiral arms \citep[e.g.][]{sellwoodSpiralInstabilitiesProvoked1984,minchevRadialHeatingGalactic2006} and bars \citep[e.g.][]{sahaEFFECTBARSTRANSIENT2010,grandVerticalDiscHeating2016}. 

Particularly in lower-mass galaxies, a variety of stellar feedback processes are thought to produce halo-like populations: shocked, outflowing gas can form new stars which inherit the underlying gas kinematics to produce kinematically hot populations \citep[e.g.][]{stinsonFeedbackFormationDwarf2009,el-badryBREATHINGFIREHOW2016}, and repeated inflow/outflow cycles due to bursty star formation can drive fluctuations in the global gravitational potential which heat stellar orbits and drive net outward migration of disk stars \citep[e.g.][]{stinsonFeedbackFormationDwarf2009,maxwellBUILDINGSTELLARHALO2012,el-badryBREATHINGFIREHOW2016}. Simulations and observations of stellar halos around massive galaxies suggest that each of these different formation mechanisms leaves distinct imprints on the properties -- including mean age, metallicity, and kinematics \citep[e.g.][]{mccarthyGlobalStructureKinematics2012,tisseraStellarHaloesSimulated2013,cooperFormationSituStellar2015,el-badryBREATHINGFIREHOW2016,kado-fongSituOriginsDwarf2022} -- of the resultant “halo” population. 

Triangulum (M33), M31’s most massive satellite \citep[$M_*\sim4.8\times10^9$~M$_\odot$:][]{corbelliDynamicalSignaturesLCDMhalo2014} presents an ideal case-study for these various halo formation mechanisms. It is sufficiently massive such that it should have experienced a number of mergers \citep[e.g.][]{conseliceStructuresDistantGalaxies2008} which could produce a stellar halo either directly from accreted debris, or associated disk heating. However, as a dwarf galaxy, it also falls into the regime where stellar feedback is expected to contribute significantly to the formation of a halo population \citep{kado-fongSituOriginsDwarf2022}. Additionally, it possesses both spiral arms and recent results from a large HST survey strongly suggest the presence of a weak ($\sim1$~kpc) bar structure \citep{lazzariniPanchromaticHubbleAndromeda2022,smercinaM33STRUCTURE2023} which could potentially contribute to disk heating and subsequent halo formation. At a distance of 859~kpc \citep{degrijsCLUSTERINGLOCALGROUP2014}, M33 is sufficiently close to study in detail. It is also relatively isolated, being located $\sim$230~kpc from M31 and thought to be on its first infall \citep{patelOrbitsMassiveSatellite2017, vandermarelFirstGaiaDynamics2019}; this suggests its halo should be relatively pristine\footnote{though the presence of an extended S-shaped stellar substructure in its outskirts \citep{mcconnachiePhotometricPropertiesVast2010} suggests it has experienced at least some dynamical perturbations in the past.}. This is critical as interactions with the $\sim10\times$ more massive M31 could easily perturb or even strip its extended stellar halo \citep{mcconnachieRemnantsGalaxyFormation2009,chapmanDynamicsSatelliteSystem2013}. 

Photometric searches for M33’s stellar halo have produced mixed results. Initial analysis of Pan-Andromeda Archaeological Survey \citep[PAndAS:][]{mcconnachieLargescaleStructureHalo2018} photometry by \citet{cockcroftUnearthingFoundationsCosmic2013} claimed tentative evidence for a halo population comprised of red giant branch (RGB) stars with scale length $\sim$20~kpc, but reanalysis of the data by \citet{mcmonigalElusiveStellarHalo2016} found no clear evidence of an extended halo. \citet{fergusonResolvingStellarOutskirts2007} find a break in M33’s radial density profile at $\sim$36’, which is now known to be accompanied by a reversal in M33’s age gradient \citep[e.g.][]{barkerStellarPopulationsOuter2007,williamsDETECTIONINSIDEOUTDISK2009}; this has been interpreted as a transition from a disk to halo component \citep[e.g.][]{barkerStarFormationHistory2011, fergusonResolvingStellarOutskirts2007}. \citet{grossiStellarStructuresOuter2011} find the distribution of an extended RGB and red clump population in individual fields at large radii is better described using an exponential profile -- suggestive of an extended disk -- rather than a power-law profile suggestive of an accreted halo \citep[e.g.][]{bullockTracingGalaxyFormation2005}. A population of old, metal-poor RR Lyrae stars \citep{sarajediniRRLyraeVariables2006, yangRRLYRAEVARIABLES2010,pritzlProbingM33Halo2011} have been suggested as belonging to a halo component, with their radial density profile similar to that of the MW and M31 stellar halos \citep{tanakulRRLyraeVariables2017}. In contrast, a lack of PNe -- which have ages up to 10~Gyr -- outside M33’s optical disk suggests a significant accreted halo component is not present \citep{galera-rosilloDeepNarrowbandSurvey2018}. 

Kinematical searches for M33’s halo have generally been more positive, although until very recently there were no studies at the same large scale as the aforementioned photometric searches. \citet{mcconnachieStellarHaloOuter2006} and \citet{hoodKinematicsChemicalAbundances2012} both identified the presence of a potential halo component with dispersion $\sim$50~km~s$^{-1}$ through analysis of line-of-sight (LOS) velocities for RGB stars, though only for a small number of stars ($<15$) located in isolated spectroscopic fields. Distinct LOS kinematics for old M33 stellar clusters has been interpreted as evidence for a kinematically hot halo population \citep[e.g.][]{chandarKinematicsStarClusters2002}, although this has also been suggested to signify only a thick disk population \citep{beasleyEvidenceTemporalEvolution2015}. 

The ongoing TRiangulum EXtended Survey \citep[TREX:][henceforth referred to as Q22]{Q22} -- which provides spectra for $>7000$ stars across M33 -- offers the ideal opportunity to comprehensively probe the presence and properties of M33’s halo. Using TREX data, \citet[][henceforth referred to as G22]{K22} modelled the kinematics of $\sim$1700 RGB stars, finding strong evidence for a significant ($\sim$22\% of the total population) kinematically hot component. Intriguingly, while this component was found to be non-rotating relative to M33's gaseous disk -- as expected for an accreted halo -- they found no significant global differences in the photometric metallicity of their halo and disk populations: more consistent with an in-situ origin. 

In order to investigate this possibility further, in this paper, we use TREX data to investigate the kinematics of an expanded sample of M33 RGB stars, as well as intermediate-age asymptotic giant branch (AGB) and carbon stars. The presence (or lack of) similarly kinematically hot components in these younger stellar populations provides critical evidence for whether in-situ mechanisms contribute significantly to the formation of M33’s halo. We present the data used for our analysis in Section~\ref{sec:data}, and outline the kinematical models used to describe the resulting velocity distributions in Section~\ref{sec:maths}. Section~\ref{sec:results} describes the results from our model fits to both old and intermediate-age stellar populations, and Section~\ref{sec:implications} discusses the associated implications for the origins of M33's halo. We summarize and conclude in Section~\ref{sec:concs}.

\section{Data} \label{sec:data}
\begin{figure}
\includegraphics[width=\columnwidth]{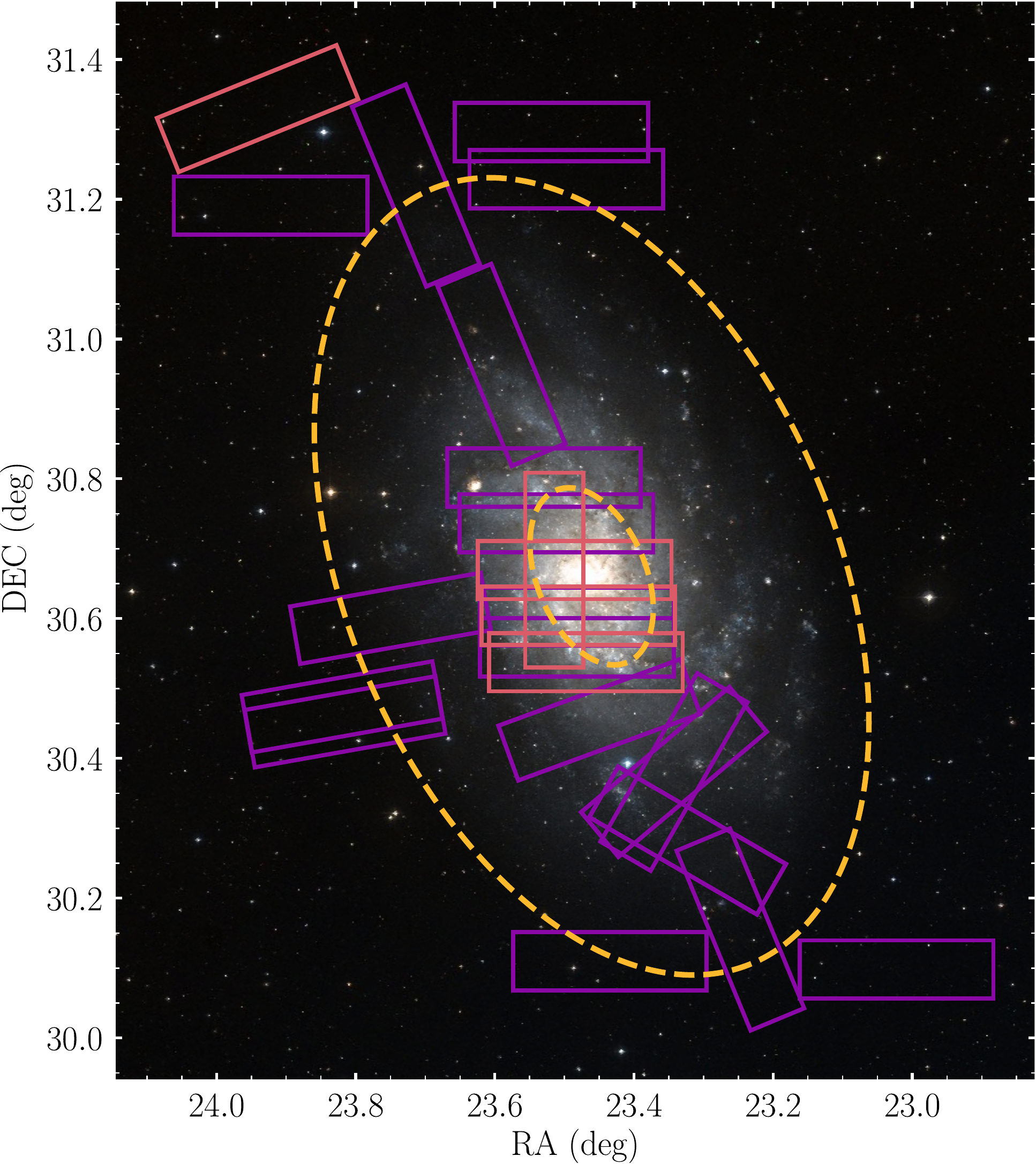}
\caption{TREX survey footprint, overlaid on a DSS color image of M33. Each rectangle approximates the shape of a DEIMOS slitmask, with purple outlines indicating TREX coverage pre-2021 as analyzed in \citetalias{K22} and \citetalias{Q22}, and red outlines indicating new masks additionally used in this analysis. Dashed orange ellipses indicate the break in M33’s surface brightness profile at $\sim$36’ \citep{fergusonResolvingStellarOutskirts2007}, and the break radius of $\sim8$' in \citet{smercinaM33STRUCTURE2023}'s model of M33's RGB halo density.}
\label{fig:maskpos}
\end{figure}

The survey overview for TREX is presented in \citetalias{Q22}, and contains details of the target selection procedures, observation characteristics, and data reduction process. Here we briefly summarize salient details. As TREX is an ongoing survey, we note that several additional masks have been observed since the release of \citetalias{Q22}, several of which are designed to increase coverage of M33’s central regions. Table~\ref{tab:masks} presents the details of additional observations used in this work, cf.\ Table 1 of \citetalias{Q22}, including mask names, positions and orientations, exposure time, number of stars observed, primary target type, and associated gratings. For these masks, isolated targets were selected primarily using Panchromatic Hubble Andromeda Treasury: Triangulum Extended Region \citep[PHATTER:][]{williamsPanchromaticHubbleAndromeda2021} photometry as occurred for TREX masks observed in 2018, described in section 2.2.1 of \citetalias{Q22}. For masks extending beyond the PHATTER footprint, this was supplemented by PAndAS photometry \citep{mcconnachieLargescaleStructureHalo2018} following the procedure for TREX masks observed in 2020. Mask pTdup1 targeted only stars with existing observations, with the aim of better constraining velocity uncertainties using repeat observations in the future (we describe our current use of repeat measurements below). Existing TREX masks have used a combination of archival HST and MegaCam/CFHT data, in addition to PHATTER and PAndAS photometry for target selection. Note that due to crowding in the very central regions of M33, we do not observe any stars in M33's bar, the extent of which is only $\sim$1~kpc \citep{smercinaM33STRUCTURE2023}. Fig.~\ref{fig:maskpos} presents the location of the new TREX masks included in this work (solid red boxes) relative to the existing TREX coverage (solid purple boxes). 

\begin{deluxetable*}{lcccccccc} \label{tab:masks}
\tablecaption{TREX masks observed in 2021}
\tablehead{\colhead{Mask Name} & \colhead{Center (J2000)} & \colhead{Mask PA} & \colhead{Exposure Time} & \colhead{N targets} & \colhead{Year} & \colhead{Grating} & \colhead{Photometry Source} & \colhead{Primary Targets} \\ 
	\colhead{} & \colhead{(h:m:s d:m:s)} & \colhead{(deg)} & \colhead{(min)} & \colhead{} & \colhead{} & \colhead{(l/mm)} & \colhead{} & \colhead{} } 

\startdata
pTA1 & 1:33:52.6 +30:32:15.9 & 90 & 78 & 198 & 2021 & 1200G & PHATTER+PAndAS & AGB+RGB \\
pTA2 & 1:33:52.6 +30:32:15.9 & 90 & 73 & 192 & 2021 & 1200G & PHATTER+PAndAS & AGB+RGB \\
pTB1 & 1:33:55.2 +30:36:14.4 & 90 & 78 & 200 & 2021 & 1200G & PHATTER+PAndAS & AGB+RGB \\
pTB2 & 1:33:55.2 +30:36:14.4 & 90 & 76 & 194 & 2021 & 1200G & PHATTER+PAndAS & AGB+RGB \\
pTC1 & 1:33:56.5 +30:40:10.4 & 90 & 77 & 193 & 2021 & 1200G & PHATTER+PAndAS & AGB+RGB \\
pTC2 & 1:33:56.5 +30:40:10.4 & 90 & 77 & 188 & 2021 & 1200G & PHATTER+PAndAS & AGB+RGB \\
pTdup1 & 1:34:03.5 +30:40:10.4 & 0 & 72 & 144 & 2021 & 600ZD & PHATTER+CFHT & range \\
pTN6 & 1:35:45.9 +31:19:46.7 & 112 & 82 & 98 & 2021 & 1200G & PAndAS & RGB \\
\enddata
\tablecomments{The position angle of the slitmask is the direction of the long axis of the slitmask measured east of north.}
\end{deluxetable*}

All masks were observed using the DEIMOS slit spectrograph \citep{faberDEIMOSSpectrographKeck2003} on the Keck II 10~m telescope. For masks targeting a mix of spectral types, the 600 line/mm grating (R$\sim$2000) is used with a central wavelength of 7200\AA, providing wavelength coverage from $\sim$4600-9800\AA. For masks primarily targeting “old” RGB stars, the 1200 line/mm grating (R$\sim$6000) is used with a central wavelength of 7800\AA, providing wavelength coverage of $\sim$6300-9800\AA. 

Spectra were reduced using the \textsc{spec2d} and \textsc{spec1d} pipelines \citep{cooperAstrophysicsSourceCode2012, newmanDEEP2GALAXYREDSHIFT2013}, which include flat-fielding, sky subtraction, extraction of 1D spectra, and measurement of the line-of-sight velocity for each object via cross-correlation against template spectra \citep{simonKinematicsUltraFaint2007}. The software \textsc{zspec} \citep{newmanDEEP2GALAXYREDSHIFT2013} was used to assign a quality measure to each star, indicating the reliability of the assigned velocity; we subsequently convert all velocities to the heliocentric frame. We account for possible mis-centering of stars within each slit using the observed position of the atmospheric A-band absorption feature (at $\sim$7600\AA) relative to night-sky emission lines. We describe the systematic variation of this “A-band correction” as a function of slit position on the mask using a polynomial function unique to each mask, fit using the A-band velocities for stars with the highest quality measure and subsequently applied to all stars on the mask; we refer to the resulting values as the “statistical” A-band correction. Velocity uncertainties for each star were estimated by summing in quadrature the uncertainty estimated by the velocity cross-correlation routine, and systematic uncertainties of 5.6~km~s$^{-1}$ and 2.2~km~s$^{-1}$ for stars observed with the 600 line/mm and 1200 line/mm gratings respectively \citep{collinsScatterUniversalDwarf2011,simonKinematicsUltraFaint2007}. We do not explicitly incorporate the additional uncertainty associated with the fitted A-band correction as the corrections themselves are generally small ($\sim3$~km s$^{-1}$) and thus this is not a dominant source of uncertainty. As we are interested in studying the kinematics of M33 stars, from this sample we remove objects with extended emission features (i.e.\ background galaxies), as well as stars with heliocentric velocities outside the range $-500\leq V_{\text{helio}}$(km~s$^{-1}$)$\leq50$, or for which the statistical A-band correction is $>3\sigma$ (where $\sigma$ is parameterized as the standard deviation) from the median statistical A-band correction for the associated mask (indicating an unreliable velocity correction). For stars with repeat observations (i.e.\ those included on mask pTdup1), if each observation is of comparable quality\footnote{i.e.\ equal quality measures as assigned by \textsc{zspec} and S/N ratios within a factor of 4} we take the average of the derived heliocentric velocities, weighted by the inverse velocity uncertainty of each observation, as the final velocity of the target. If one observation is of significantly higher quality, we take only the velocity derived from that observation as the final velocity of the target.

\begin{figure*}
\includegraphics[width=\textwidth]{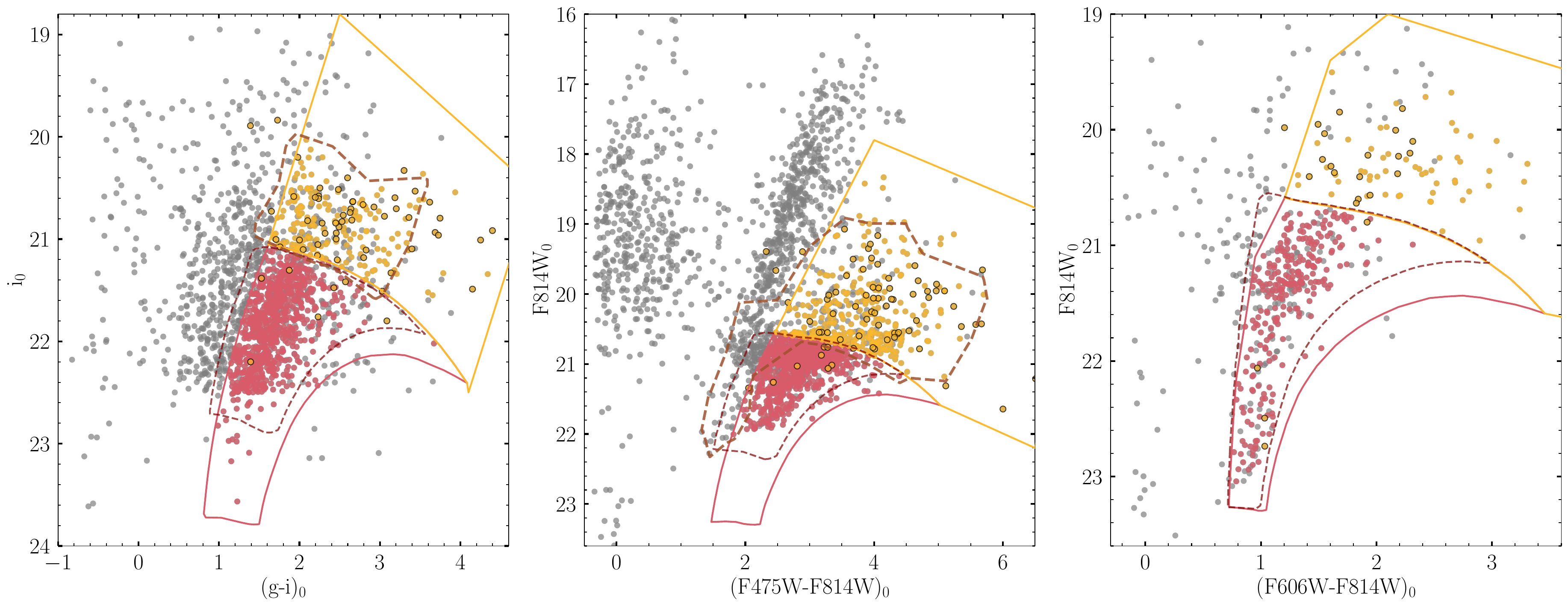}
\caption{Extinction-corrected color-magnitude diagrams of targeted stars with LOS velocity measurements (grey). The left panel shows stars with CFHT/MegaCam photometry; predominantly from PAndAS but also including a very small fraction of other archival data. The center panel shows stars with HST photometry in the F457W and F814W bands; predominantly from PHATTER but also including some other archival data. The right panel shows stars with archival HST photometry in the F606W and F814W bands. Solid lines show the selection boxes used to isolate RGB and AGB stars, with darker dashed lines indicating the RGB and AGB selections of \citetalias{K22} and \citetalias{Q22} respectively. While there is large overlap between these selections, both the \citetalias{K22} and \citetalias{Q22} samples include younger, bluer stars. Red and orange points indicate RGB and AGB stars included in our final selection; orange points with black outlines indicate spectroscopically-identified carbon stars included in our “intermediate-age” sample.}
\label{fig:cmd}
\end{figure*}

Guided by Padova isochrones \citep{bressanPARSECStellarTracks2012}\footnote{version 3.6, accessed at \url{http://stev.oapd.inaf.it/cgi-bin/cmd}} of age 10~Gyr (matching those used by \citetalias{K22}), and setting M33's distance modulus to 24.67 \citep{degrijsCLUSTERINGLOCALGROUP2014}, we use optical color-magnitude selection boxes to classify the remaining stars into different broad age groups. Fig.~\ref{fig:cmd} shows the selection boxes used to define “old” RGB (red) and “intermediate-age” AGB (orange) star samples in each of the filter sets from which targets are selected, overlaid the full sample of stars for which velocities are measured. Within these selections, we re-classify stars with spectral features indicating they are either carbon stars or weak-CN stars (see Appendix A of \citetalias{Q22}): regardless of their position on the CMD, carbon stars are associated with the “intermediate-age” sample, and weak-CN stars are associated with a “young” sample not discussed further in this paper. For reference, in Fig.~\ref{fig:cmd} we also show the CMD selections for \citetalias{K22}'s RGB sample (dashed maroon lines) and \citetalias{Q22}'s AGB sample (dashed brown lines); we discuss differences between these samples below. 

While infrared (IR) color-magnitude selections are typically used to distinguish between AGB and RGB stars \citep[e.g.][]{williamsPanchromaticHubbleAndromeda2021,smercinaM33STRUCTURE2023}, we utilize optical photometry only in order to maintain consistent selections across the entire survey region: IR photometry is only available for stars within the PHATTER region. We test the efficiency of our selection by comparing the nominal AGB sample selected using our optical criteria within the PHATTER region to that selected using the IR criteria in Table 1 of \citet[][henceforth referred to as S23]{smercinaM33STRUCTURE2023} over the same region. We find the selections are largely consistent; $\sim78$\% of the AGB sample selected using the optical criteria are also classified as AGB stars using the IR criteria, with the remainder being either classified as RGB stars ($\sim10$\%), or outside the bounds of the selection area in either a gap between \citetalias{smercinaM33STRUCTURE2023}'s AGB and RGB selections ($\sim$7\%), or blueward of \citetalias{smercinaM33STRUCTURE2023}'s AGB selection boundary ($\sim5$\%). There are 10 additional stars we nominally associate with the RGB sample ($<1$\%) which are associated with the IR-selected AGB sample. 

We also remove likely MW contaminants via visual inspection by excluding stars with the surface-gravity-sensitive \ion{Na}{1} doublet (at $\sim$8190\AA) present in their spectra \citep{gilbertNewMethodIsolating2006}; we exclude 64 nominal AGB and 59 nominal RGB stars using this criterion. While this is effective at identifying dwarf stars in the MW disk (see Appendix A of \citetalias{K22}), it is not useful for identifying distant main sequence turnoff (MSTO) stars in the MW halo. We quantify the level of MW contamination we expect in our remaining sample by comparison with the Besan\c{c}on MW model \citep{robinSyntheticViewStructure2003}\footnote{accessed at \url{https://model.obs-besancon.fr/}}. Within our RGB selection region, $\sim$27\% of MW model stars have distances $>$6~kpc -- our nominal cutoff between MW “disk” and “halo” stars when looking in the direction of M33. Assuming the 59 stars in our RGB selection region with \ion{Na}{1} absorption therefore represent $\sim$73\% of the MW contaminants along the line of sight, we expect on order of 22 unidentified MW halo stars in our final RGB sample ($\sim$1\% of the total). In contrast, within our AGB selection region, there are zero MW model stars exceeding the distance cutoff, suggesting that there are effectively no unidentified MW halo stars remaining in this sample. 

\begin{figure*}
\includegraphics[width=0.75\textheight]{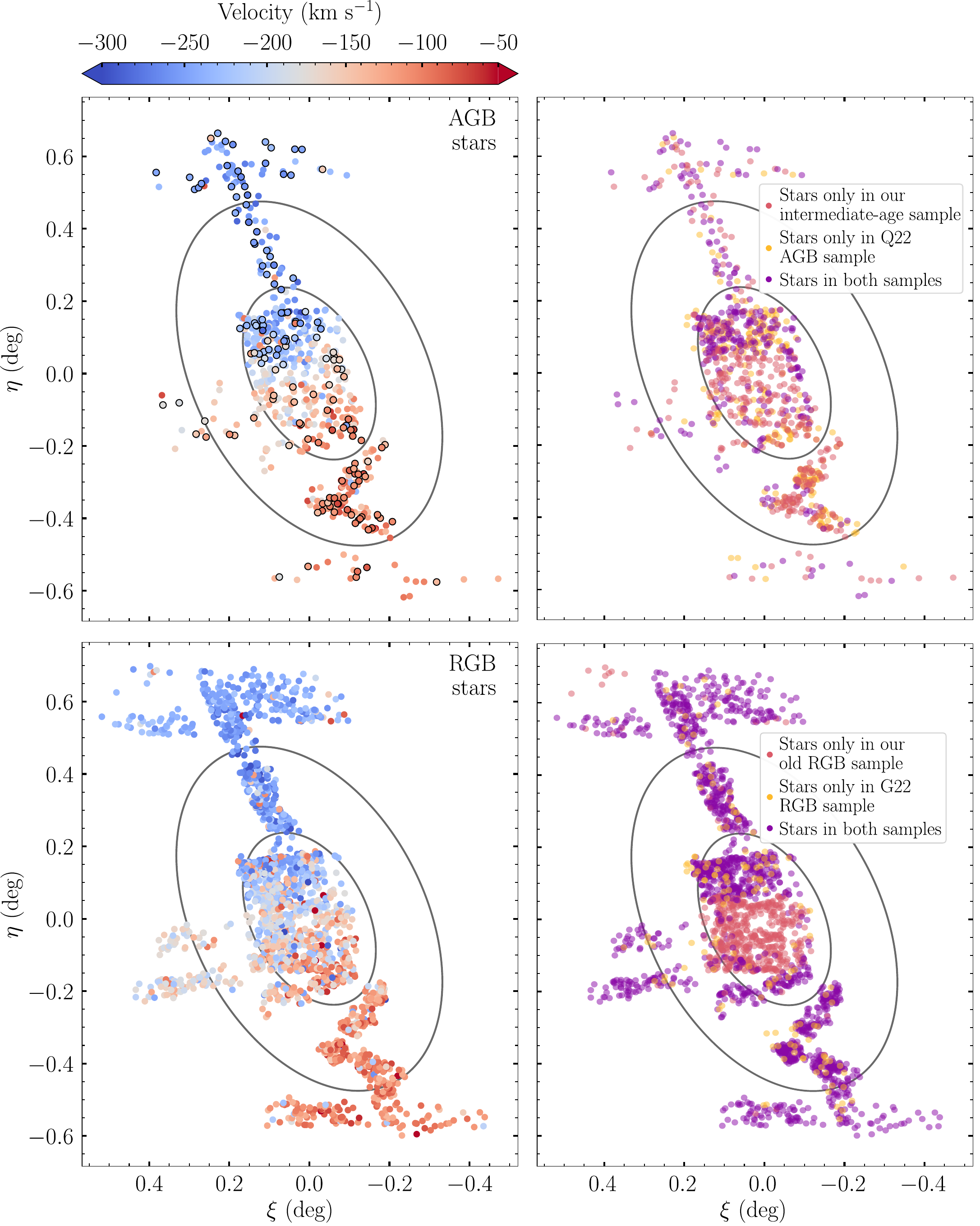} 
\caption{On-sky positions of our “intermediate-age” AGB (top) and “old” RGB (bottom) stellar samples. In the left panels, all points are color-coded by their heliocentric LOS velocity. Points with black outlines indicate spectroscopically-identified carbon stars. The velocity gradient associated with M33's rotation is clearly visible. In the right panels, points are colored by if a star is present in only our sample (red), only in the \citetalias{Q22} or \citetalias{K22} samples (yellow), or both (purple). There are significant differences in our intermediate-age sample compared to that of \citetalias{Q22}, discussed in \S\ref{sec:data}. The solid grey ellipses indicate projected radii in the plane of M33's disk of 15’ and 30’, which we use in \S\ref{sec:spatial} to identify radial variations in M33's kinematic properties.}
\label{fig:map}
\end{figure*}

The final sample of stars analyzed in this paper contains 2032 “old” RGB stars, and 671 “intermediate-age” (516 AGB and 155 carbon) stars. The left panels of Fig.~\ref{fig:map} plots the spatial distribution of these stars on-sky, color-coded by their (heliocentric) LOS velocities. We also compare our old and intermediate-age populations to those described in \citetalias{K22} and \citetalias{Q22} respectively in the right panels of Fig.~\ref{fig:map}.

Our old RGB sample has significant overlap with that of \citetalias{K22}, and increases their sample size by 22\%. The majority of additional stars included our sample are from new TREX masks observed since the publication of \citetalias{K22}: predominantly filling significant gaps in the previous spatial coverage of RGB stars in M33's inner southern disk, but also adding to the sample in M33's far northeast (pTN6). Small differences in the selected stars from existing TREX masks reflect a differing blue cutoff between our RGB sample and that of \citetalias{K22} as seen in Fig.~\ref{fig:cmd}: \citetalias{K22} includes all stars redder than the [Fe/H]$=-2.32$ isochrone, while we select only stars redder than isochrones of [Fe/H]$=-2.2$ for the archival HST photometry and [Fe/H]$=-1.8$ for PAndAS and PHATTER photometry. The stricter cutoff in the latter case is designed to reduce contamination from younger red helium-burning or AGB stars, which do abut and overlap the blue edge of the RGB (see e.g.\ Figure 2 of \citetalias{Q22}), but this does consequently exclude the most metal-poor RGB stars included in \citetalias{K22}'s sample. We also use a mildly more generous red (i.e. metal-rich) cutoff of [Fe/H]$=-0.2$ compared to \citetalias{K22}, but functionally this only increases our sample size by $\sim$15 stars. 

In contrast, there is much less overlap between our intermediate-age sample, and an analogous sample of AGB stars from \citetalias{Q22}. While there are similarly additional stars from the new TREX masks included in our sample, there are also key differences in the selection regions used to define AGB stars between the two samples. The AGB selection in \citetalias{Q22} incorporates full 6-filter PHATTER photometry where this is available to identify AGB stars located blueward of the RGB; these are not included in our sample as we only utilize optical photometry, and as discussed above there is significant overlap between stars of different ages in this region of the optical CMD. In addition, in regions where archival HST photometry is available, \citetalias{Q22} only include spectroscopically-identified carbon stars in their intermediate-age sample; our selection includes additional AGB stars identified purely photometrically. 
The main effect this has on the resulting sample is a difference in their mean age (derived from isochrone fitting of targets within the PHATTER region as in \citetalias{smercinaM33STRUCTURE2023}, assuming a constant star-formation history): our intermediate-age sample is $\sim3$~Gyr old (with a full range of ages $\sim0.2-5.1$~Gyr), while the average AGB age in \citetalias{Q22} is $\sim1$~Gyr (including stars of ages $\sim0.56-2.2$~Gyr). 

\section{Stellar velocity models} \label{sec:maths}
Motivated by the analysis in \citetalias{K22}, we model the stellar LOS velocity distribution for our two samples relative to that of M33’s \ion{H}{1} disk, following the procedure outlined in section 3.1 of \citetalias{K22} and detailed below in \S\ref{sec:h1maths}. This inherently assumes that the stars are all rotating in the same thin plane as the \ion{H}{1} disk, and are on circular orbits; we discuss the implications of this in Section~\ref{sec:implications}.  We test two potential models for the stellar velocity distribution which allow for either co- or counter-rotation relative to the \ion{H}{1} disk (\S\ref{sec:rotmod}) or systemic offsets between the stellar and gaseous velocity distributions (\S\ref{sec:offmodel}); Section~\ref{sec:compmod} describes how we compare between these models.

\subsection{\ion{H}{1} Velocity model}\label{sec:h1maths}
The reference \ion{H}{1} disk model we utilize is the tilted ring model of \citet[][henceforth referred to as K17]{kamKinematicsMassDistribution2017}. Their Table 4 describes the rotation speed ($V_{\text{HI,rot}}$), inclination ($i$), and position angle ($\text{PA}$) measured east of north to the receding semi-major axis of the \ion{H}{1} disk, each as a function of deprojected angular distance from M33's center in the disk plane ($R_{\text{disk}}$). We interpolate within this model, assuming a series of infinitesimally thin rings, each with its own inclination, position angle, and rotation speed. 

For a given star $j$, we computed an initial estimate of $R_{\text{disk}}$ for each star based on global values for M33's disk inclination and PA ($52^\circ$ and $202^\circ$ respectively: \citetalias{kamKinematicsMassDistribution2017}) using Eq.~\ref{eq:rdisk}, and then revised the $R_{\text{disk}}$ estimate using the inclination and PA values from the interpolated \citetalias{kamKinematicsMassDistribution2017} model at the star's initial $R_{\text{disk}}$ estimate. We iterated calculation of $R_{\text{disk}}$ until the change in inclination and PA from the previous and current estimates was $<0.01^\circ$ and $<0.35^\circ$ respectively; this corresponds to 10\% of the typical resolution of inclination and position angle changes in Table 4 of \citetalias{kamKinematicsMassDistribution2017}. 

\begin{equation}\label{eq:rdisk}
R_{\text{disk},j} = \sqrt{\gamma_j^2 + \left(\frac{\beta_j}{\cos(i_j)}\right)^2}
\end{equation}

where 
\begin{eqnarray}\label{eq:alphas}
\gamma_j &= \eta_j\cos\left(\text{PA}_j\right) + \xi_j\sin\left(\text{PA}_j\right) \\
\beta_j &= \xi_j\cos\left(\text{PA}_j\right)-\eta_j\sin\left(\text{PA}_j\right)
\end{eqnarray}

Here, $\xi_j$ and $\eta_j$ are the M33-centered sky coordinates of star $j$, calculated from its on-sky coordinates ($\alpha_j$,$\delta_j$) using Eq.~\ref{eq:skypos}; we take the center coordinates of M33 ($\alpha_0$,$\delta_0$) as 01h33m50.9s, +30d39m36s to match those used in \citetalias{K22}.

\begin{eqnarray}\label{eq:skypos}
\xi_j &= \frac{\cos(\delta_j)\sin(\alpha_j-\alpha_0)}{\sin(\delta_j)\sin(\delta_0)+\cos(\delta_j)\cos(\delta_0)\cos(\alpha_j-\alpha_0)} \\
\eta_j &= \frac{\sin(\delta_j)\cos(\delta_0)-\cos(\delta_j)\sin(\delta_0)\cos(\alpha_j-\alpha_0)}{\sin(\delta_j)\sin(\delta_0)+\cos(\delta_j)\cos(\delta_0)\cos(\alpha_j-\alpha_0)} 
\end{eqnarray}

The LOS component of the \ion{H}{1} rotation velocity at the position of a given star, $V_{\text{HI LOS},j}$, is then calculated using Eq.~\ref{eq:gasvel}:
\begin{equation}\label{eq:gasvel}
V_{\text{HI LOS},j} = V_{\text{HI rot},j} \times \cos{\left(\theta_j\right)} \sin{\left(i_j\right)} 
\end{equation}

where $V_{\text{HI rot},j}$ is the rotation speed of the \ion{H}{1} disk at $R_{\text{disk},j}$, and $\theta_j$ is the azimuthal angle of star $j$ in the plane of the \ion{H}{1} disk, measured relative to the semi-major axis, as calculated in Eq.~\ref{eq:theta}.

\begin{equation}\label{eq:theta}
\theta_j = \frac{\beta_j}{\gamma_j\cos\left(i_j\right)}
\end{equation}

We define $V_{\text{offset},j}$ as the difference between the \ion{H}{1} LOS velocity at the position of a given star ($V_{\text{HI LOS},j}$) and the observed LOS velocity of each star per Eq.~\ref{eq:voff}. The sign convention for $V_{\text{offset},j}$ is such that a lag in the rotation velocities of the stars relative to the \ion{H}{1} model results in a negative $V_{\text{offset},j}$ in the northeastern (approaching) side of the disk, and a positive $V_{\text{offset},j}$ in the southwestern (receding) side of the disk. A stellar population which rotates in the plane of the \ion{H}{1} disk and with no mean offset from the \ion{H}{1} LOS velocity at any given location $j$ will therefore have a $V_{\text{offset}}$ distribution centered at the velocity of the gas disk (i.e.\ $\langle V_{\text{offset}}\rangle=0$~km~s$^{-1}$)

\begin{equation}\label{eq:voff}
V_{\text{offset},j} = V_{\text{HI LOS},j} - V_{\text{LOS},j}
\end{equation}

Given findings in \citetalias{K22} that RGB stars do rotate more slowly than the measured \ion{H}{1} rotation speed, we use the formalism of \citetalias{K22} and model the LOS velocities of the stars $V_{\text{modLOS},j}$ as a fraction ($f_{\text{rot}}$) of the rotation speed of the \ion{H}{1} disk model, as in Eq.~\ref{eq:vfrot}. We parameterize the stellar rotation as a fraction of the \ion{H}{1} disk rotation, rather than an explicit asymmetric drift value, to account for the fact that our samples span a large range of radii ($0.25\leq R_{\text{disk}}\text{ (kpc)}\leq17.7$), over which the \ion{H}{1} rotation curve of M33 increases significantly (from $\sim$40 to 125~km~s$^{-1}$: \citetalias{kamKinematicsMassDistribution2017}).

\begin{equation}\label{eq:vfrot}
V_{\text{modLOS},j} = v_{\text{sys}} + f_{\text{rot}}\times V_{\text{HI LOS},j}
\end{equation}

Here, $v_{\text{sys}}$ is the systemic velocity of M33's disk ($-180$~km~s$^{-1}$: \citetalias{kamKinematicsMassDistribution2017}).

For each star, we then calculate the difference $V_{\text{offmod},j}$ between the observed LOS velocity of each star $V_{\text{LOS},j}$ and its modelled LOS velocity using Eq.~\ref{eq:voffrot}, using the same sign convention as for $V_{\text{offset},j}$. 

\begin{equation}\label{eq:voffrot}
V_{\text{offmod},j} = V_{\text{modLOS},j} - V_{\text{LOS},j}
\end{equation}

A stellar population which rotates in the plane of the \ion{H}{1} disk at some fraction of the \ion{H}{1} rotation velocity will therefore have a $V_{\text{offmod}}$ distribution centered at zero, with some width $\sigma$; we assume this distribution is Gaussian. 

\subsection{Rotating population models}\label{sec:rotmod}
Motivated by findings in \citetalias{K22} that more than one Gaussian distribution in $V_{\text{offmod}}$ is required to describe their RGB sample, we first choose a flexible model which describes the $V_{\text{offmod}}$ distribution by a mixture model of $N$ Gaussian components, each with dispersion $\sigma_N$ and relative rotation speed $f_{\text{rot},N}$. Accordingly, the likelihood of the model for an individual star is given by Eq.~\ref{eq:likelihood}. 

\begin{equation}\label{eq:likelihood}
\mathcal{L}_j = \sum_N f_N\times\mathcal{N}(V_{\text{offset},j}(f_{\text{rot},N})|0,\sigma_{N})
\end{equation}

Here, $\mathcal{N}(V_j|0,\sigma)$ denotes a normalized Gaussian centered at the systemic velocity of the gas disk (i.e.\ with mean $\mu=0$) and dispersion $\sigma$, evaluated at the velocity of star $j$ ($V_j$) in the specified frame of reference (in our parameterization, $V_{\text{offmod},N}$ for each component, calculated using $f_{\text{rot},N}$ for that component). $f_N$ is the fraction of the total number of stars present in the $N$th component, such that $\sum_N f_N=1$. For a two-component model, this is identical to Equation 9 of \citetalias{K22}. We refer to this formalism as the “rotating model”, since it inherently assumes that each Gaussian component is rotating in the plane of the gas disk. However, the model parameter $f_{\text{rot}}$ provides the flexibility for a component to be best-fit as a non-rotating component ($f_{\text{rot}}$=0), or a counter-rotating component ($f_{\text{rot}}<0$), in the plane of the gas disk.

We note that we do not explicitly incorporate the velocity uncertainties for each star in the model, and $\sigma$ is therefore the observed, not intrinsic, velocity dispersion of each component. An approximation of the intrinsic dispersion can be calculated as $\sigma_{N,\text{int}}^2=\sigma_{N,\text{obs}}^2-\delta_s^2$, where $\delta_s$ is the median velocity uncertainty for all stars contributing to the fit. For the intermediate-age sample, $\delta_s\sim5.7$~km~s$^{-1}$, while for the old RGB sample, $\delta_s\sim4.3$~km~s$^{-1}$. 

To determine the best-fitting parameters for each of the Gaussian components within the mixture model, we sample the posterior distribution of the model parameters -- i.e.\ $f_N$\footnote{Note that since the total fraction of stars in each component must sum to one, it is only necessary to fit $N-1$ values of $f_N$.},$f_{\text{rot},N}$, and $\sigma_{N}$, collectively referred to as $\Theta$ -- using the nested sampling package \textsc{dynesty} \citep{speagleDynestyDynamicNested2020} in order to maximize the overall log-likelihood for the model, $\sum_j\log\mathcal{L}_j$. We utilize uniform priors for all model parameters. Parameter values subsequently quoted in this work are the 50th percentile of the marginalized posterior probability distribution, with the quoted uncertainties corresponding to the 16th and 84th percentiles.

The probability $P_{j,N}$ of an individual star $j$ being associated with the $N$th component is therefore given by Eq.~\ref{eq:probability}, evaluated at the best-fitting model parameter values $\Theta$. 
\begin{equation}\label{eq:probability}
P_{j,N}=\frac{f_N\times\mathcal{N}(V_{\text{offset},j}(f_{\text{rot},N})|0,\sigma_{N})}{\mathcal{L}(\Theta)_j}
\end{equation}

\subsection{An offset model}\label{sec:offmodel}
The above “rotating model” formalism naturally describes the velocity distribution of stellar populations with kinematics tied to those of the gas disk; including for example `thin' and `thick' stellar disks, and is also likely a reasonable description of any halo component formed in-situ via e.g.\ dynamical heating. However, it is not necessarily the most appropriate model to describe the kinematics of an accreted halo population. Such a component may not rotate in the plane of the \ion{H}{1} disk, or at all. 

In addition, interactions can introduce significant shifts in the position and velocity of a galaxy's center of mass (CoM) on short timescales \citep[e.g.][]{gomezItMovesDangers2015}. This can introduce apparent velocity offsets between disk stars (which, with relatively short dynamical timescales, move with the CoM) and distant halo stars \citep[which, with relatively long dynamical timescales, are slow to respond to the shift in CoM: e.g.][]{petersenReflexMotionMilky2020}. This so-called “reflex motion” has been observed in the Milky Way due to the infall of the LMC, with an apparent systemic velocity offset of $\sim$30~km~s$^{-1}$ for halo stars relative to the MW disk \citep{petersenDetectionMilkyWay2021}. Given that M33 displays indications of having undergone interactions in the past, with a significant stellar and gaseous warp in its outskirts \citep[e.g.][]{mcconnachiePhotometricPropertiesVast2010, putmanDISRUPTIONFUELINGM332009}, it is not implausible that a similar reflex motion could be observed here. 

Accordingly, we test fitting a two-component model which can describe this scenario, which we refer to as the “offset model”. While this model describes the velocity distribution of the “disk” in the same way as the “rotating” model, the velocity distribution of the “halo” component is described as a single Gaussian component in LOS space directly (\textit{not} in $V_{\text{offset}}$ space), with mean $\mu_{\text{halo}}$ and dispersion $\sigma_{\text{halo}}$. Eq.~\ref{eq:offlikelihood} describes the likelihood of this model. 

\begin{eqnarray}\label{eq:offlikelihood}
\mathcal{L} = \sum_j(1-f_{\text{halo}})\times\mathcal{N}(V_{\text{offset,disk},j}(f_{\text{rot,disk}})|0,\sigma_{\text{disk}}) \nonumber \\ +f_{\text{halo}}\times\mathcal{N}(V_{\text{LOS},j}|\mu_{\text{halo}},\sigma_{\text{halo}}) 
\end{eqnarray}

The probability of a individual star belonging to either the `disk' or `halo' component can be calculated analogously to the “rotating model” case per Eq.~\ref{eq:probability}. 

\subsection{Comparing models}\label{sec:compmod}
As we test several “rotating” model configurations with varying numbers of Gaussian components (discussed further in Section \ref{sec:results}), and the “offset” model (discussed in Section~\ref{sec:offset}). , we require a way compare the resulting model fits. To do so, for each model configuration we estimate the marginal likelihood $\mathcal{Z}$ of the model given our data, often referred to as the evidence, per Eq.~\ref{eq:evidence} using the nested sampling implementation \textsc{dynesty}. Here, $\pi(\Theta)$ represents the priors associated with the model parameters. 

\begin{equation}\label{eq:evidence}
\mathcal{Z}=\int\mathcal{L}(\Theta)\pi(\Theta) d\Theta
\end{equation}

We compare model configurations by calculating the difference in log-evidence ($\Delta\log_{10}\mathcal{Z}$, abbreviated henceforth as “DLZ”) for different model configurations, related to the Bayes factor \citep[$K$:][]{jeffreysTheoryProbability1998} per Eq.~\ref{eq:bayes}. 

\begin{eqnarray}\label{eq:bayes}
K_{12} = \frac{\mathcal{Z}_1}{\mathcal{Z}_2}  \\
\log_{10}K_{12} = \log_{10}\mathcal{Z}_1 - \log_{10}\mathcal{Z}_2 = \Delta\log_{10}\mathcal{Z} \nonumber 
\end{eqnarray}

We interpret the relative strength of the evidence for preferring one model configuration over another using the interpretation of \citet{kassBayesFactors1995}, as in Table~\ref{tab:bayes}. 

\begin{deluxetable*}{cccl} \label{tab:bayes}
\tablecaption{Bayes factor thresholds used to interpret model evidence. Adapted from \citet{kassBayesFactors1995}.}
\tablehead{$2\ln(K_{12})$ range & $K_{12}$ range &$\Delta\log_{10}\mathcal{Z}$ range & Description} 

\startdata
$\leq0$ & $\leq1$ & $\leq0$ & Model 1 not preferred over model 2 \\
$0<x<2$ & $1<x<3$ & $0<x<0.5$ & Evidence for model 1 barely worth mentioning \\
$2\leq x<6$ & $3\leq x<20$ & $0.5\leq x<1.3$ & Positive/substantial evidence preferring model 1 \\
$6\leq x<10$ & $20\leq x<150$ & $1.3\leq x<2.2$ & Strong evidence preferring model 1 \\
$\geq10$ & $\geq150$ & $\geq2.2$ & Decisive evidence preferring model 1
\enddata
\end{deluxetable*}

\section{High-dispersion populations in M33}\label{sec:results}
Motivated by the results of \citetalias{K22}, we initially fit the “rotating” model to our data (\S\ref{sec:rotres}) for both the old and intermediate-age populations, and discuss spatial variations in the resulting fits in \S\ref{sec:spatial}. We then discuss results from fitting the “offset” model to our data in \S\ref{sec:offset}.

\subsection{Rotating models}\label{sec:rotres}
Given \citetalias{K22} find differences in the number of “rotating” model components preferred for populations of varying ages in M33, we test fitting one, two, and three Gaussian components (each centered at $V_{\text{offset}}=0$) in the framework of our “rotating” model. The one-component model (for which $f_N=1$, and hence $f_{\text{rot, disk}}$ and $\sigma_{\text{disk}}$ are the only parameters fit) matches that which best describes the young star sample of \citetalias{K22}. The two-component model (which, like the three-component model, fits $f_N$, $f_{\text{rot,}N}$ and $\sigma_{\text{disk,}N}$ for each component, under the condition that $\sum_Nf_N=1$) matches that which best describes their RGB sample. 

\begin{figure*}
\includegraphics[width=0.5\textwidth]{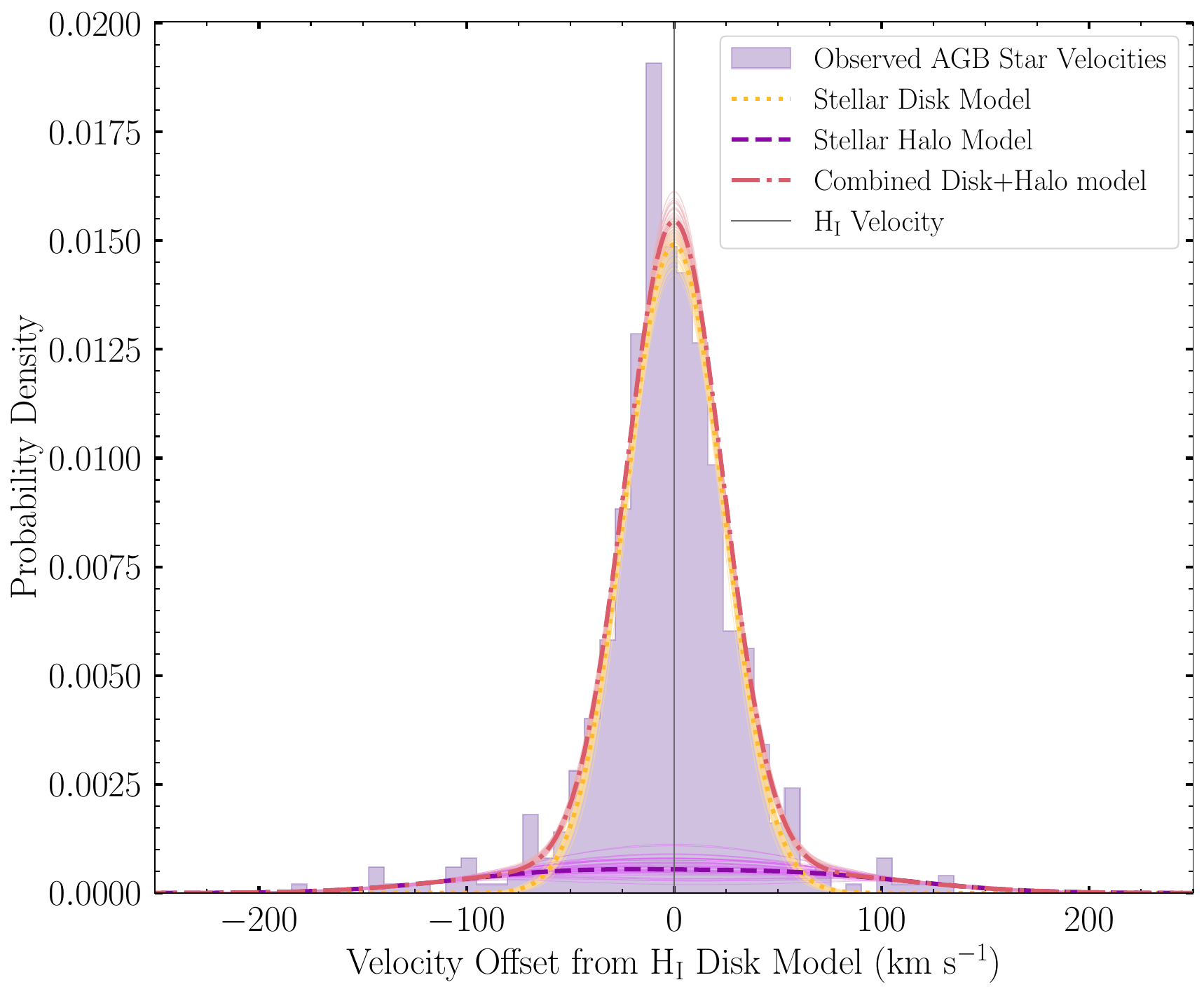}\includegraphics[width=0.5\textwidth]{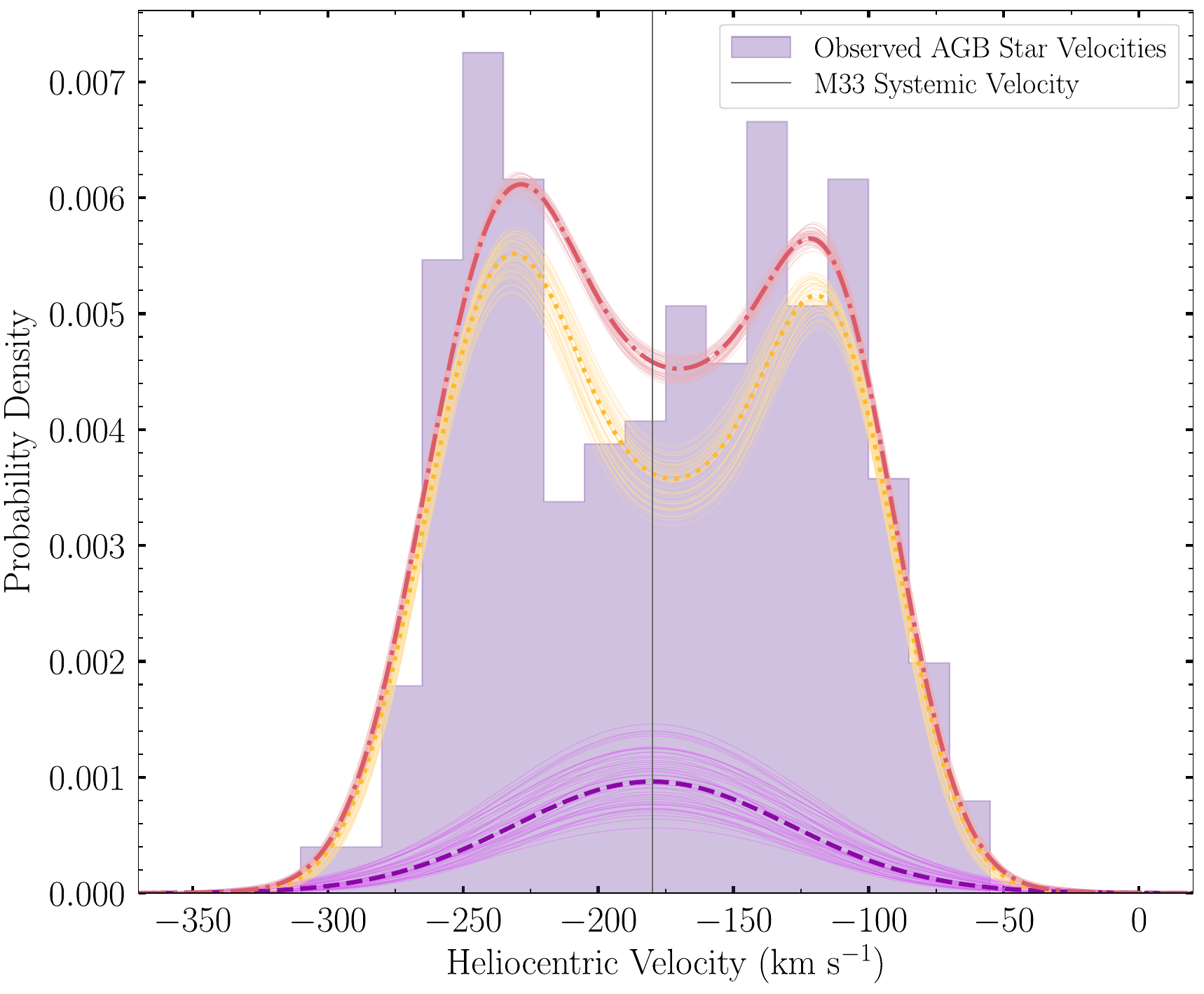}
\includegraphics[width=0.5\textwidth]{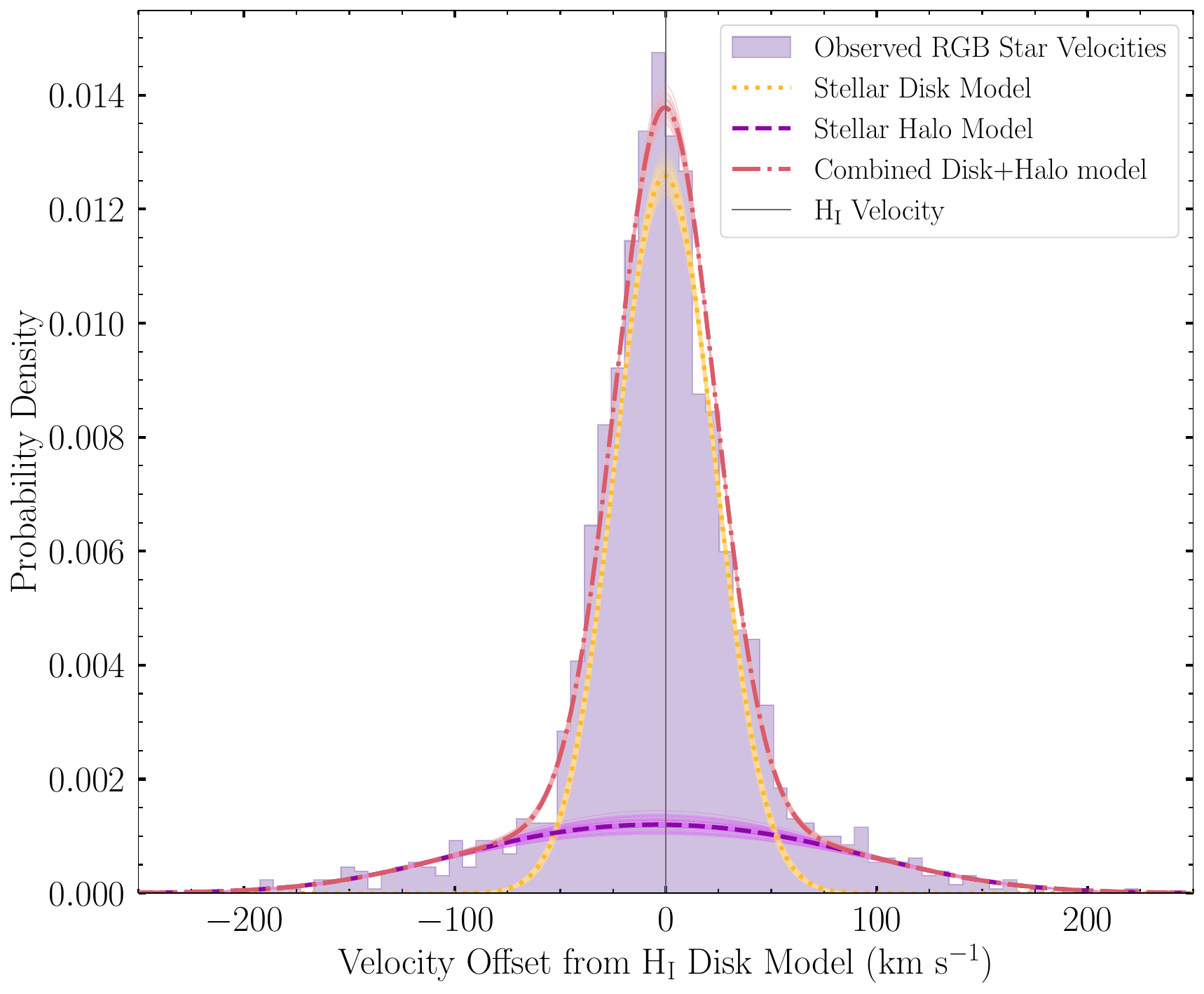}\includegraphics[width=0.5\textwidth]{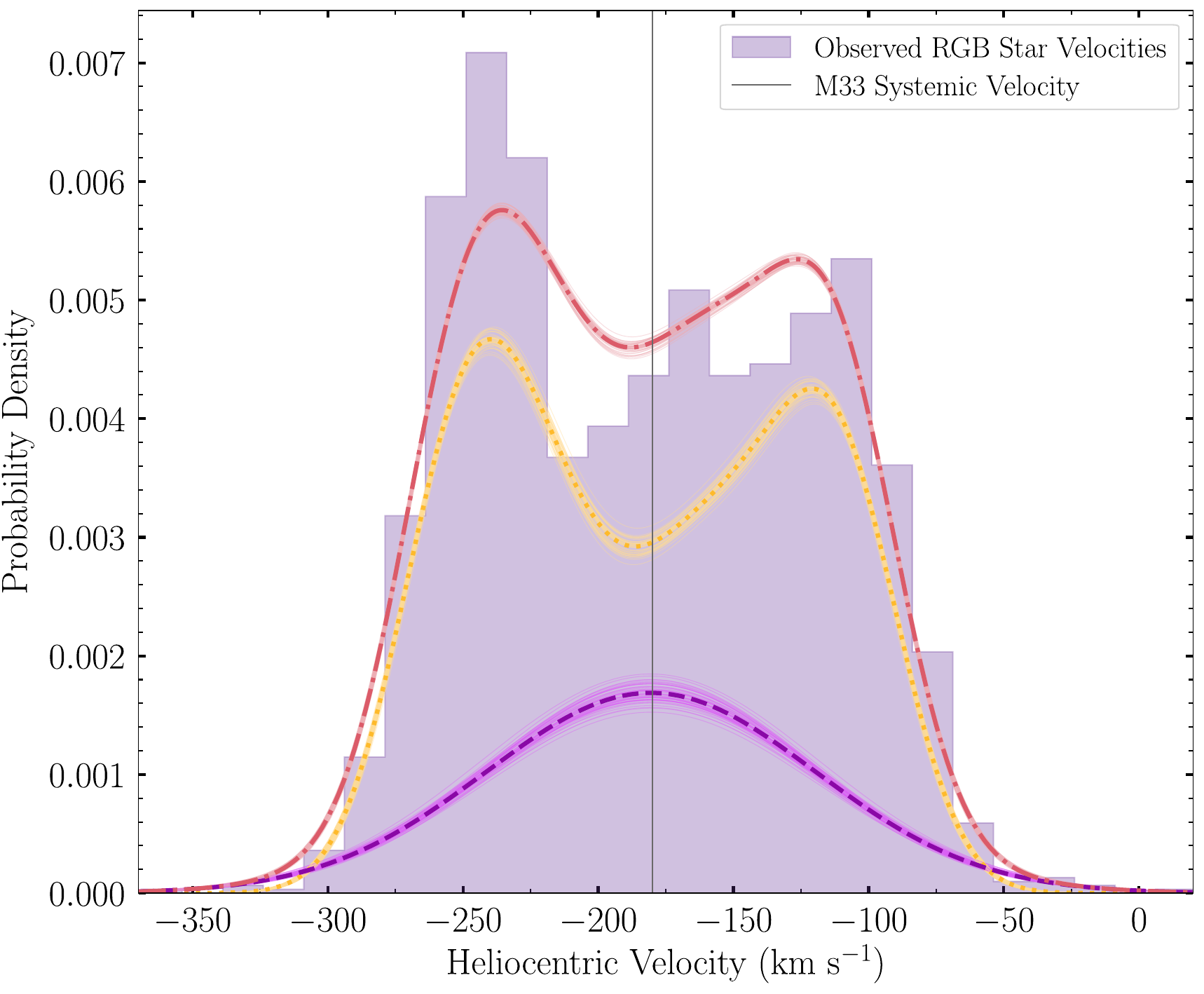}

\caption{Distribution of the stellar LOS velocities (right) and offset velocities from the predicted \citetalias{kamKinematicsMassDistribution2017} \ion{H}{1} gas velocities ($V_{\text{offset}}$:left) for the full “intermediate-age” sample (top) and “old” RGB sample (bottom). Dashed lines indicate the best-fitting two-component “rotating” model (red), as well as the individual disk (yellow) and high-dispersion “halo” components (magenta). Fifty random draws from within the 1$\sigma$ uncertainties on the best-fitting model parameters are shown as thin solid lines for each component. M33's systemic velocity is shown as a solid grey line. Both old and intermediate-age populations prefer a halo component which minimally co-rotates in the plane of the gas disk, with the relative halo fraction for the intermediate-age component significantly lower than that for the old population.}
\label{fig:rotres}
\end{figure*}

Fig.~\ref{fig:rotres} shows the heliocentric LOS velocity distribution (right), plotted on the left as the difference between the predicted \ion{H}{1} velocity and the observed stellar velocity (i.e.\ $V_{\text{offset}}$), for the intermediate-age (top) and old (bottom) populations, along with the best-fitting two-component models. For both populations, there are clearly stars with large velocity offsets from those predicted by the \ion{H}{1} disk model at their location, indicating a single-component fit is insufficient to describe the stars. This is borne out when comparing the evidence for these two model configurations, with the two-component model “decisively preferred” (DLZ$\sim$130 for the old stars, and $\sim$38.5 for the intermediate age stars: see Table~\ref{tab:bayes}) over the single-component model. This remains true when the intermediate-age sample is split, and the spectroscopically-identified carbon stars (DLZ$\sim$10.5) and photometrically-selected AGB stars (DLZ$\sim$27) are fit independently. A three-component model is not preferred relative to the two-component model for the intermediate-age stars, and only has evidence “barely worth mentioning” for the old stars; consequently, we discuss further only the results for the two-component model. We subsequently refer to the lower- and higher-dispersion components of this model as the “disk” and “halo” components respectively. 

Tables~\ref{tab:agbrot} and \ref{tab:rgbrot} present the best-fitting two-component model parameters and associated uncertainties for the intermediate-age and old populations respectively. Our results for the old stars are entirely consistent with results from \citetalias{K22} for their RGB star sample; as our old stellar sample largely matches that from \citetalias{K22}, with an increased number of stars simply from new observations, this is expected. 

\begin{deluxetable*}{@{\extracolsep{4pt}}cccccccc} \label{tab:agbrot}
\tablecaption{Best-fitting parameters for the “rotating” two-component model of the intermediate-age population, which allows each component to rotate at a fraction of the \ion{H}{1} disk velocity.}
\tablehead{\colhead{}&\colhead{} & \colhead{} & \multicolumn{2}{c}{Disk parameters} & \multicolumn{3}{c}{Halo parameters}\\ \cline{4-5} \cline{6-8}
	\colhead{Radial range} & \colhead{Spatial range} & \colhead{$N_{\text{stars}}$} & \colhead{$f_{\text{rot, disk}}$} & \colhead{$\sigma_{\text{disk}}$} & \colhead{$f_{\text{halo}}$} & \colhead{$\sigma_{\text{halo}}$} & \colhead{$f_{\text{rot, halo}}$}}

\startdata
& All & 671 & $0.89^{+0.02}_{-0.02}$ & $23.1^{+1.0}_{-1.1}$ & $0.10^{+0.04}_{-0.03}$ & $52.9^{+7.1}_{-8.1}$ & $-0.21^{+0.31}_{-0.3}$ \\
All & NE & 338 & $0.90^{+0.02}_{-0.02}$ & $21.3^{+1.4}_{-2.2}$ & $0.12^{+0.06}_{-0.03}$ & $43.7^{+17.3}_{-10.1}$ & $-0.62^{+0.67}_{-0.22}$ \\
& SW & 333 & $0.90^{+0.03}_{-0.03}$ & $24.3^{+1.4}_{-1.4}$ & $0.11^{+0.06}_{-0.04}$ & $51.0^{+9.0}_{-7.3}$ & $0.18^{+0.26}_{-0.33}$ \\ \hline
$<15$ & All & 391 & $0.96^{+0.04}_{-0.04}$ & $25.3^{+1.4}_{-1.8}$ & $0.11^{+0.07}_{-0.04}$ & $36.4^{+9.2}_{-7.2}$ & $-0.48^{+0.51}_{-0.34}$ \\
$15-30$ & All & 193 & $0.88^{+0.02}_{-0.02}$ & $21.7^{+1.4}_{-1.3}$ & $0.08^{+0.04}_{-0.03}$ & $46.5^{+14.4}_{-12.3}$ &$ \-0.19^{+0.35}_{-0.27}$ \\
$30+$ & All & 87 & $0.84^{+0.02}_{-0.02}$ & $15.3^{+1.4}_{-1.3}$ & $0.10^{+0.04}_{-0.03}$ & $77.8^{+14.1}_{-14.9}$ & $-0.50^{+0.45}_{-0.42}$ 
\enddata
\tablecomments{Columns give the fraction of the \ion{H}{1} rotation speed at which the disk and halo are rotating ($f_{\text{rot, disk}}$, $f_{\text{rot, halo}}$; negative values correspond to counter-rotation in the plane of the gas disk); disk and halo velocity dispersions ($\sigma_{\text{disk}}$, $\sigma_{\text{halo}}$ in km~s$^{-1}$); and the total fraction of the population in the halo component ($f_{\text{halo}}$). Spatial divisions of the sample are made as described in \S\ref{sec:spatial}.}
\end{deluxetable*}

\begin{deluxetable*}{@{\extracolsep{4pt}}cccccccc} \label{tab:rgbrot}
\tablecaption{Best-fitting parameters for the “rotating” two-component model of the old (RGB) population, which allows each component to rotate at a fraction of the \ion{H}{1} disk velocity.}
\tablehead{\colhead{}&\colhead{} & \colhead{} & \multicolumn{2}{c}{Disk parameters} & \multicolumn{3}{c}{Halo parameters}\\ \cline{4-5} \cline{6-8}
	\colhead{Radial range} & \colhead{Spatial range} & \colhead{$N_{\text{stars}}$} & \colhead{$f_{\text{rot, disk}}$} & \colhead{$\sigma_{\text{disk}}$} & \colhead{$f_{\text{halo}}$} & \colhead{$\sigma_{\text{halo}}$} & \colhead{$f_{\text{rot, halo}}$}}

\startdata
& All & 2032 & $0.88^{+0.01}_{-0.01}$ & $22.2^{+0.7}_{-0.7}$ & $0.25^{+0.02}_{-0.02}$ & $61.2^{+2.2}_{-2.1}$ & $0.12^{+0.08}_{-0.09}$ \\
All & NE & 1028 & $0.86^{+0.01}_{-0.01}$ & $21.2^{+0.9}_{-0.9}$ & $0.22^{+0.03}_{-0.03}$ & $64.8^{+3.5}_{-3.3}$ & $-0.02^{+0.12}_{-0.13}$ \\
& SW & 1004 & $0.90^{+0.02}_{-0.02}$ & $23.0^{+1.3}_{-1.2}$ & $0.31^{+0.04}_{-0.04}$ & $56.5^{+2.8}_{-2.6}$ & $0.27^{+0.09}_{-0.12}$ \\ \hline 
& All & 912 & $1.00^{+0.03}_{-0.03}$ & $27.3^{+1.6}_{-1.6}$ & $0.34^{+0.05}_{-0.05}$ & $48.7^{+3.6}_{-3.7}$ & $-0.33^{+0.17}_{-0.16}$ \\
$<15$ & NE & 509 & $0.93^{+0.04}_{-0.04}$ & $27.0^{+1.4}_{-1.4}$ & $0.26^{+0.04}_{-0.03}$ & $44.8^{+4.6}_{-4.0}$ & $-0.60^{+0.16}_{-0.14}$ \\
& SW & 403 & $1.17^{+0.05}_{-0.05}$ & $22.0^{+3.3}_{-2.9}$ & $0.54^{+0.07}_{-0.07}$ & $51.6^{+2.9}_{-2.9}$ & $0.09^{+0.13}_{-0.16}$ \\ \hline
& All & 605 & $0.88^{+0.01}_{-0.01}$ & $20.9^{+1.1}_{-1.0}$ & $0.18^{+0.04}_{-0.03}$ & $57.0^{+4.9}_{-4.6}$ & $0.20^{+0.13}_{-0.16}$ \\
$15-30$ & NE & 215 & $0.91^{+0.02}_{-0.02}$ & $19.7^{+1.4}_{-1.4}$ & $0.15^{+0.05}_{-0.04}$ & $57.0^{+9.8}_{-8.7}$ & $0.13^{+0.22}_{-0.26}$ \\
& SW & 390 & $0.86^{+0.02}_{-0.02}$ & $21.9^{+1.5}_{-1.4}$ & $0.19^{+0.05}_{-0.05}$ & $57.5^{+6.3}_{-5.9}$ & $0.19^{+0.16}_{-0.23}$ \\ \hline
& All & 515 & $0.84^{+0.01}_{-0.01}$ & $18.9^{+0.8}_{-0.7}$ & $0.11^{+0.02}_{-0.02}$ &$ 76.6^{+8.5}_{-7.3}$ & $-0.03^{+0.18}_{-0.19}$ \\
$30+$ & NE & 304 & $0.81^{+0.01}_{-0.01}$ & $16.8^{+0.8}_{-0.7}$ & $0.09^{+0.02}_{-0.02}$ & $84.8^{+9.2}_{-9.4}$ & $-0.24^{+0.23}_{-0.23}$ \\
& SW & 211 & $0.91^{+0.03}_{-0.03}$ & $23.2^{+1.5}_{-1.5}$ & $0.09^{+0.04}_{-0.03}$ & $31.7^{+18.7}_{-6.1}$ & $-0.37^{+0.44}_{-0.17}$
\enddata
\end{deluxetable*}

\subsubsection{Disk properties}
We find the best-fit parameters for old and intermediate-age disk populations mostly agree within uncertainty. This behavior for the disk dispersion ($\sigma_{\text{disk}}$) agrees with measurements of local dispersion for M33 disk stars in \citetalias{Q22}, which find that this is relatively constant with age. Our best-fit disk dispersion values for both old and intermediate-age populations are, however, $\sim$7~km~s$^{-1}$ higher than those in \citetalias{Q22} for similar age populations. This is likely driven at least partially by the increased number of stars at smaller radii -- which have a higher dispersion than those at larger radii (see \S\ref{sec:spatial}) -- in our samples relative to those of \citetalias{Q22}, and the different selections used to define the intermediate-age samples for which $\sigma_{\text{disk}}$) is measured. In addition, as the values reported by \citetalias{Q22} are averages of local dispersion measurements for disk stars (defined as those with $P_{\text{disk}}>$80\% based on the \citetalias{K22} velocity model), rather than kinematical fits to the overall disk dispersion as obtained from our method, it is not unexpected that slightly different values are measured. The fact that $f_{\text{rot,disk}}$ is consistent for both old and intermediate-age stars is also somewhat different to findings from \citetalias{Q22} that the mean asymmetric drift for intermediate-age stars is less than that for older RGB stars. However, as our intermediate-age sample is on average older (and thus closer in age to the RGB sample) than that of \citetalias{Q22}, a more consistent $f_{\text{rot,disk}}$ is not unexpected. In addition, as our parameterization is quite different from that of \citetalias{Q22} -- who measure the average local asymmetric drift for disk stars -- these measurements are not directly comparable. 

We note that the dispersion values we report are the \textit{observed} LOS dispersions. Projection effects due to M33's inclination and the finite thickness of its disk \citep[expected to be $\sim200$~pc based on scaling relations:][]{williamsHighresolutionRadiativeTransfer2019} act to inflate the observed LOS dispersion, as do the uncertainties in the measured LOS velocities of individual stars. To estimate the magnitude of the former effect, we use a toy model of M33's disk, similar to that described in \citet{guhathakurtaKinematicsM31Thick2021} and \citet{quModelingStellarKinematics2022}. In the model, we generate an exponential distribution (set by M33's scale length) of stars on circular orbits in the plane of M33's disk (or a fixed vertical distance above it set by the input scale height). The model is then rotated to match the apparent orientation of M33 on the sky, and the corresponding LOS velocity of each star calculated. We measure the local LOS dispersion due to the projection of stars at overlapping radii at any point on the sky by taking the standard deviation of the LOS velocity distribution of stars within 1' of the specified point. We find the magnitude of the dispersion due to this geometric effect decreases with projected disk radius. For radii within 15' (i.e.\ encompassing the majority of our sample), this effect increases the observed LOS dispersion by $\sim3.5$~km~s$^{-1}$; comparable to that associated with the mean velocity uncertainty of the stars (see Section~\ref{sec:rotmod}). Accounting for both effects gives an approximate intrinsic LOS dispersion of $\sim$21~km~s$^{-1}$ for both the old and intermediate age populations. 

Nonetheless, our measured dispersion is similar to the velocity dispersion of old stars in the extreme outer disk of the Large Magellanic Cloud \citep[$\sim20$~km~s$^{-1}$:][]{C20}, and only $\sim$50\% of that in the inner LMC disk \citep[$\sim40$~km~s$^{-1}$: e.g.][]{vasilievInternalDynamicsLarge2018}. As disk velocity dispersion typically increases with galaxy mass \citep{bottemaStellarKinematicsGalactic1993}, and M33 is $\sim$1.7$\times$ more massive than the LMC \citep{vandermarelLargeMagellanicCloud2006}, at first glance it might be expected that M33 would have a larger dispersion than the LMC, contrary to that observed. However, the LMC is an actively interacting system, having just passed pericenter on its infall to the MW \citep{beslaAreMagellanicClouds2007} and recently experienced a close interaction with the Small Magellanic Cloud \citep[][]{zivickProperMotionField2018,choiRecentLMCSMC2022} -- both of which act to increase the velocity dispersion of its disk \citep{C22, choiRecentLMCSMC2022}. Given evidence for M33 having experienced such recent strong interactions is lacking, it is therefore perhaps unsurprising that its disk dispersion is lower than that of the LMC at comparable radii. 

\subsubsection{Halo properties}\label{sec:haloprops}
The best-fit parameters for the halo component are also largely consistent for the two stellar populations. While the intermediate-age stellar halo has a dispersion $\sim8$~km~s$^{-1}$ lower than that of the old stellar halo, this is still consistent within uncertainty. As in \citetalias{K22}, we find the old halo component has very slight prograde rotation ($f_{\text{rot,halo}}=0.12$), but this is consistent with zero at $1.3\sigma$. The fractional rotation of the intermediate-age stellar halo is, however, relatively poorly constrained, and consistent with zero well within uncertainty. The main difference between the old and intermediate-age populations is the relative contribution of the halo component ($f_{\text{halo}}$): $\sim$10\% of the intermediate-age population is associated with the halo, while this is $\sim$25\% for the old population. This is a $>3\sigma$ difference. 

\begin{figure*}
\includegraphics[height=6.3cm]{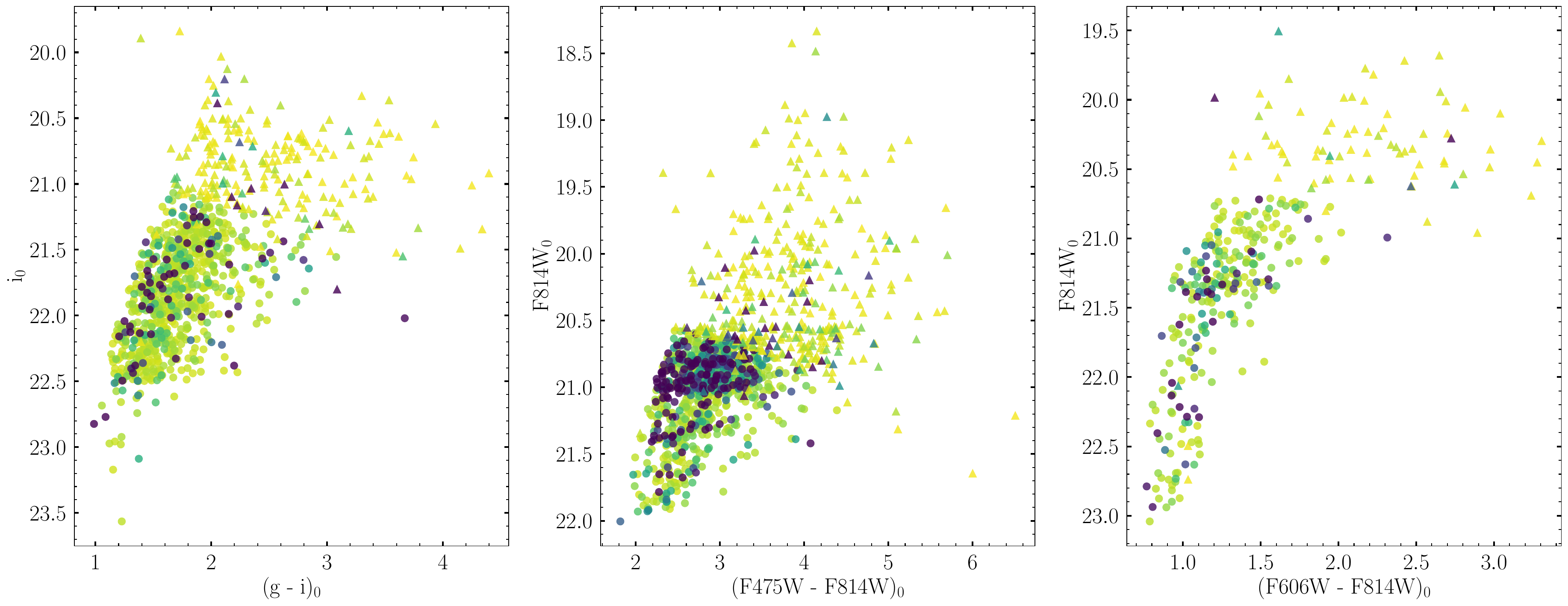} 
\raisebox{12pt}{\includegraphics[height=6cm]{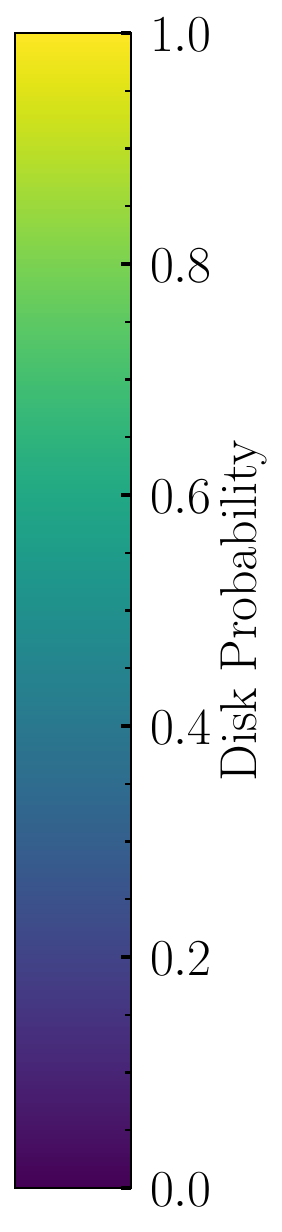}} 
\caption{CMD distributions for our intermediate-age (triangular) and old (circular points) stellar populations, color-coded by their probability of belonging to the low-dispersion disk-like component in a two-component fit. Dark purplish points therefore indicate points which are likely associated with M33's high-dispersion halo component; these are distributed approximately evenly across each CMD given the underlying target selection function.}
\label{fig:cmdlike}
\end{figure*}

In order to verify that the strong preference for the two component model in the intermediate-age sample is not being driven by potential contamination from older RGB populations, we additionally test making a number of adjustments to the photometric AGB selection, and re-run our “rotating” model on the adjusted sample. We first test fitting the PHATTER IR-selected AGB sample discussed in \S\ref{sec:data} and find negligible ($<1\sigma$) differences in the resulting best-fit parameters or preference for a two-component model. This further suggests that our use of an optical-only selection for the wider intermediate-age population does not bias our results. We additionally test adjustments to our overall optical AGB selection by varying the distance modulus of M33 within its $2\sigma$ uncertainty from \citet{degrijsCLUSTERINGLOCALGROUP2014}, which is used to define the selection boxes in Fig.~\ref{fig:cmd}. The resulting most conservative adjustment produces an AGB/RGB cutoff magnitude $\sim0.14$~mag brighter than our canonical selection. We similarly find no differences in the best-fit parameters or preference for a two-component model when using this sample. Finally, we note that stars which have a high likelihood ($P_j>$80\%) of belonging to the halo component are distributed approximately randomly across the CMD (i.e.\ not correlated with color or magnitude beyond that associated with the underlying density distribution due to our varying initial TREX target selection), as seen in Fig.~\ref{fig:cmdlike}. In combination, this suggests that our intermediate-age population genuinely prefers the inclusion of a high-dispersion halo component in the model. 

In summary, we find that a two-component model which includes both a disk-like (i.e.\ low-dispersion, co-rotating with the gas disk) and halo-like (i.e.\ high-dispersion, minimal or no co-rotation relative to the gas disk) component, is strongly preferred to describe the kinematics of intermediate-age and old stellar populations across M33. While the parameters of the halo component are generally similar for the two populations, potentially indicating a similar formation mechanism (though see Section~\ref{sec:implications} for further discussion), there is a significant difference in the fraction of each population which contributes to the halo: 25\% of the old stellar population is associated with the halo, but this is only 10\% for the intermediate-age population. In general, this suggests the mean age of the halo is older than that of the disk: as expected for both an accreted (old, metal-poor) halo, or one formed through in-situ dynamical heating processes (in which older populations have had a longer time to experience perturbations and be heated into a halo-like orbit). 

\subsection{Spatial variations}\label{sec:spatial}
\citetalias{K22} find kinematical complexity in their old RGB sample, with different model parameters preferred for different radial regions, as well as in the northern and southern halves of M33’s disk. Accordingly, we also investigate potential spatial variations in stellar kinematics by splitting our samples into three radial bins ($\leq15$’, $15-30$’, and $\geq30$’), and within each of those regions by their north/south position, parameterized by whether the \citetalias{kamKinematicsMassDistribution2017} \ion{H}{1} velocity at the location of each star is greater than (south) or less than (north) M33’s systemic velocity of $-180$~km~s$^{-1}$. The resulting best-fit parameters are also included in Tables \ref{tab:agbrot} and \ref{tab:rgbrot}. 

\begin{figure}
\includegraphics[height=0.9\textheight]{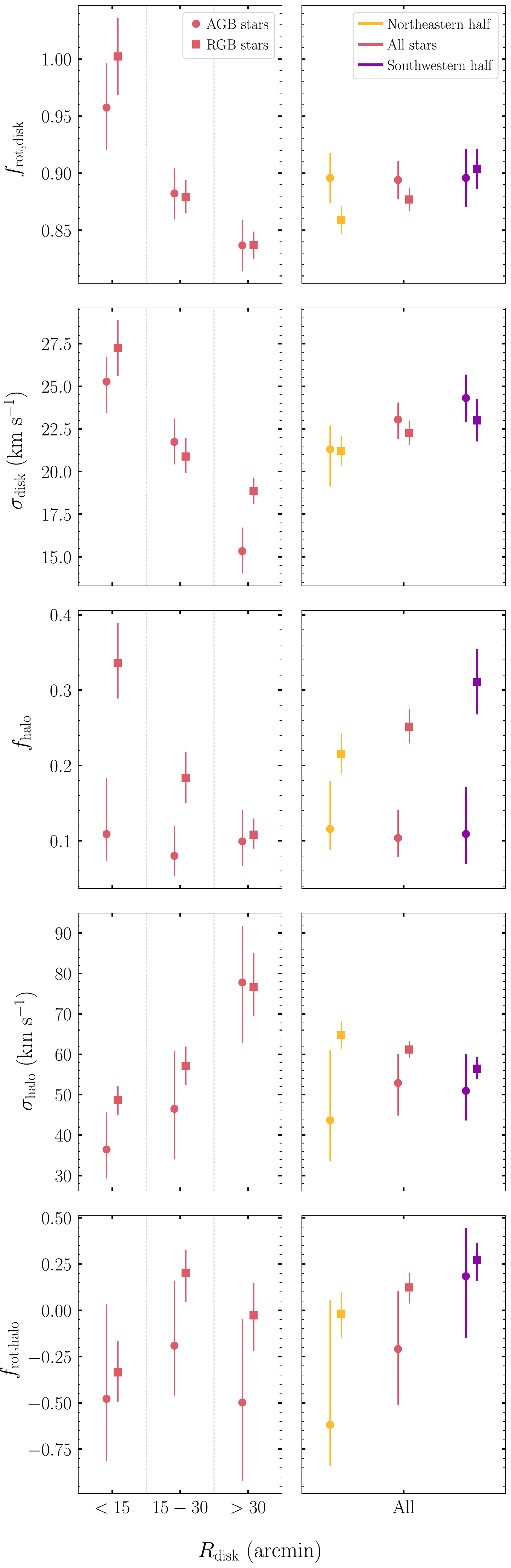}
\caption{Best-fitting model results for the two-component “rotating” model, fit to the old RGB (square points) and intermediate-age (circular points) stellar samples, split into radial bins (left) and northern/southern halves of the disk (right). Similar radial trends are observed for both populations, excepting the fraction of stars associated with the halo component ($f_{\text{halo}}$): this decreases with radius for the old population, but remains approximately constant for the intermediate-age component.}
\label{fig:comprot}
\end{figure}

We find model parameters for our old population are consistent with those from \citetalias{K22} within uncertainty for all spatial regions. The lower number of stars in our intermediate-age population prevents us from splitting this into similarly fine spatial regions while maintaining convergence of our model fits, or from independently fitting the carbon stars and photometrically-selected AGB stars in each region for comparison. We do, however, fit each of the three radial regions, and the global northern and southern regions of the disk, for the full intermediate-age sample. Fig~\ref{fig:comprot} compares the best-fitting model parameters for each of these regions for the two-component models for the old and intermediate-age populations. 

\subsubsection{Variation in velocity dispersions}\label{sec:vardisp}
The best-fitting parameters for the disk component are largely consistent between the two populations, with both the dispersion ($\sigma_{\text{disk}}$) and the fractional rotation speed ($f_{\text{rot,disk}}$) decreasing slightly with radius. We note the latter of these is not necessarily suggestive that the rotation speed of the stellar disk is deceasing with radius -- indeed, \citetalias{Q22} find that outside the initial rise, the stellar rotation curve of M33 is relatively flat\footnote{though note their derivation, as in our model, assumes the stars follow the same geometry as the \ion{H}{1} disk.} -- but may simply indicate that the stellar kinematics are less closely tied to those of the \ion{H}{1} disk at larger radii. The decrease in disk dispersion with radius, however, is likely genuine: it matches that seen in \citetalias{Q22}, who find a decease in velocity dispersion as a function of projected radius for both intermediate-age and old M33 disk stars, and is also consistent with trends of decreasing dispersion with radius observed in the similar-mass LMC \citep[e.g.][]{vasilievInternalDynamicsLarge2018,gaiacollaborationGaiaEarlyData2021a}. 

In contrast, the halo dispersion \textit{increases} with radius for both the old and intermediate-age populations. This is different to behavior observed in the more massive MW and M31 halos, where dispersion mildly decreases with radius in the innermost regions of the halo (like those probed here), and remains approximately constant at larger radii \citep[e.g.][]{battagliaRadialVelocityDispersion2005,brownVELOCITYDISPERSIONPROFILE2010,gilbertGlobalPropertiesM312018}. However, we caution over-interpretation of this trend, which is largely driven by results from the outermost radial bin. While there is still an increase in halo dispersion between the two inner radial bins, these are consistent within $1\sigma$ for the intermediate-age population, and $1.4\sigma$ for the old stellar population. In addition, there are additional kinematical complexities for the old stellar population within 15’ (discussed further below) which suggest our model may not be fully describing the halo kinematics in this region. 

We additionally note the halo dispersion for the old stellar population is $\sim$8~km~s$^{-1}$ higher in the northeastern region of the disk. However, this is also driven entirely by the outermost bin: the dispersion in M33’s northeastern outskirts ($\sim$85~km~s$^{-1}$) is $>2.5\times$ that of its southwestern outskirts. If we instead fit the northeastern and southwestern halves of the disk within 30’ only, the old halo dispersion in the northeastern half of the disk is reduced by $\sim$13~km~s$^{-1}$, and is consistent with both that of the southwestern half of the disk, and the global fit to this radial region. The halo fraction and disk parameters for the old stellar population are unchanged when fitting just this restricted radial region.

The comparatively extreme dispersions for both the intermediate-age and old populations in the outermost bin are similar to those observed in regions of the LMC outskirts strongly affected by interactions \citep{C22}; and this radial bin includes the region where M33’s \ion{H}{1} disk is becoming strongly warped, indicative of perturbation. As the majority of stars in this bin are located in the vicinity of M33’s gaseous (and stellar) warp, rather than perpendicular to it (see Fig.~\ref{fig:map}), we suggest our model may instead be attempting to fit a different population in the outermost bin, particularly in the northeastern half -- perhaps tidally disrupted disk stars perturbed in a past interaction as suggested to dominate M33’s outskirts by \citet{mcconnachiePhotometricPropertiesVast2010}. In order to determine if there is both a perturbed extended disk as well as a smooth continuation of the halo population observed at smaller radii, uniform coverage of M33’s outskirts -- particularly including further observations of regions perpendicular to M33’s stellar and gaseous warps -- is necessary. Future observations with e.g.\ the wide-field Dark Energy Spectroscopic Instrument (DESI), as recently performed across M31’s inner halo \citep{deyDESIObservationsAndromeda2023}, will be key in providing additional insight into this possibility.

\subsubsection{Variations in halo fraction}\label{sec:varfrac}
Intriguingly, the fraction of stars associated with the halo component ($f_{\text{halo}}$) is consistent with radius for the intermediate-age population, but decreases significantly -- from 34\% to 11\%, a difference of $>4\sigma$ -- for the old population. The old halo fraction also differs between the northern and southern halves of the disk, with an additional 10\% of the old population associated with the halo component in the southwest compared to the northeast. In contrast, for the intermediate-age population, there is no significant difference in the halo fraction (or any other parameter) between the northeastern and southwestern halves of the disk. 

The decrease in the old halo fraction with radius aligns with recent results from \citetalias{smercinaM33STRUCTURE2023}, who map the radial density profile of old RGB stars in M33 out to $\sim$4.5~kpc ($\sim$18.5’)\footnote{note that in comparison, our spectroscopic data extend to $>30'$.} using PHATTER and Spitzer photometry. They find this is best described by a combination of an exponential disk and a broken power-law “halo” component ($R_{\text{break}}\sim$2~kpc/8’) which dominates in the inner regions of M33. If their model is extrapolated to larger distances, the resulting halo fraction within each of the radial regions we define matches what we observe for the old population. While their “global average” halo fraction of 15.7\% within radii up to $R_{\text{disk}}=17.5$~kpc (the distance of our furthermost spectroscopic measurement) is slightly lower than the 25\% inferred from our fit to the full old RGB sample, this is expected given our spectroscopic sample is not uniformly spatially distributed, and is instead dominated by stars at small radii, for which the halo fraction is greater. 

It is worth noting that within the limited radial extent of the photometry used to derive the density profile of \citetalias{smercinaM33STRUCTURE2023}, their best-fit broken-power-law model (for which the halo fraction \textit{decreases} as a function of radius) is degenerate with a single (i.e.\ unbroken) power law, for which the halo fraction instead \textit{increases} as a function of radius. This is because the halo power-law slope is shallower in the single-power-law model than the exponential profile describing the disk density, while in the broken-power-law model, the halo power-law slope outside the break radius is steeper than the exponential disk profile. Accordingly, the relative contribution of the halo is larger and increases with radius in the single-power-law model. While such an increasing halo fraction is more consistent with predictions for the fraction of accreted (i.e.\ halo) stars in cosmological simulations \citep[e.g.][]{rodriguez-gomezStellarMassAssembly2016,davisonEAGLEViewEx2020}, as noted by \citetalias{smercinaM33STRUCTURE2023} this is not consistent with the trend inferred from our kinematical fits, and also requires an extended halo surface brightness profile above the upper limits obtained from PAndAS photometry at large radii \citep{mcmonigalElusiveStellarHalo2016}. 

\citetalias{smercinaM33STRUCTURE2023} do not derive a similar radial density profile for M33's intermediate-age populations due to the limited spatial extent of the (infrared) PHATTER photometry required to precisely identify AGB stars, and the presence of significant spiral arms compared to the relatively smooth distribution of older RGB stars. However, future precise wide-field IR photometry obtained from e.g.\ the Nancy Grace Roman telescope offers the possibility of deriving such a radial profile for intermediate-age stars, which would provide a useful point of comparison given the relatively uniform halo fraction inferred from our kinematical fits to the intermediate-age population. 

\subsubsection{Variation in halo rotation}\label{sec:varhrot}
There is no obvious trend with radius for the fractional rotation speed of the halo component ($f_{\text{rot,halo}}$) for either the old or intermediate-age populations. We do note, however, that for both populations the fractional rotation is more negative (indicative of rotation in the plane of the \ion{H}{1} disk, but in the direction opposite that of the \ion{H}{1} gas) in the northeastern half of the disk, and more positive in the southwestern half of the disk, though the best-fit values are consistent within uncertainty. In the case of the old population, this trend is driven predominantly by stars at small radii ($<$15’) in the northeastern half of the disk: the best-fit fractional rotation here is $-0.6^{+0.16}_{-0.14}$ -- inconsistent with zero at $4\sigma$ -- indicating these stars are strongly counter-rotating relative to the \ion{H}{1} gas. In comparison, over the same radial region, stars in the southwestern half of the disk have a fractional rotation for the halo consistent with zero. \citetalias{K22} found a similar result, but as their sample included $\sim$3x fewer stars in the southwestern region, this difference in kinematics between the northeastern and southwestern halves of the disk at small radii was not considered particularly robust. Our sample contains comparable numbers of stars in both halves of the inner M33 disk, and $\sim30$\% more stars than \citetalias{K22} at radii$<$15' in the northeastern half of the disk, indicating there is a genuine difference in the kinematics of old RGB stars at small radii. 

We note that the preferred value for $f_{\text{rot,halo}}$ for the intermediate-age population in the northeastern half of the disk is also strongly negative, though the large uncertainties on this parameter mean this is not statistically significant as in the case of the old population. The significantly lower number of stars in the intermediate-age sample also precludes us from directly determining if this nominal counter-rotation is similarly localized to the inner northeastern disk within 15'. However, we do note that the overall trends which result from the localized counter-rotation -- i.e., more negative $f_{\text{rot,halo}}$ values in the global northeast, and at all radii $<15$’ -- are seen in both the old population and the intermediate-age population\footnote{albeit at a less statistically significant level for the intermediate-age population.}. As such, it is possible that both intermediate-age and old halo stars in the inner northeastern disk are similarly strongly counter-rotating. Further observations of intermediate-age stars in this region are necessary to substantiate this hypothesis.  

Interestingly, \citetalias{smercinaM33STRUCTURE2023} find asymmetries in the spatial distribution of stars of all ages in the inner M33 disk, which they suggest are indicative of tidal interactions. While we see no evidence of strong asymmetries in the kinematics of \textit{disk} stars for either the old or intermediate-age populations at any radii, it seems plausible that the kinematical asymmetries we find for the \textit{halo} stars in the innermost region (see Table \ref{tab:rgbrot}) are similarly indicative of perturbations due to interactions; we discuss this idea in more detail in Section~\ref{sec:interactimp}. 

\subsection{An offset intermediate-age population?}\label{sec:offset}
As discussed in \S\ref{sec:offmodel}, while our “rotating” model should provide a reasonable description of a halo formed in-situ via disk heating mechanisms, alternate models (which do not require halo rotation and allow for systematic offsets from the systemic galaxy velocity) may provide a better description of an accreted halo. Accordingly, and motivated by the fact that $f_{\text{rot,halo}}$ is always consistent with zero within uncertainty for the intermediate-age population, we additionally test fitting a two-component “offset” model to both the old and intermediate-age populations. 

We find there is at minimum “substantial” evidence preferring this “offset” model over the “rotating” model for the intermediate-age population (DLZ$\sim$1.3). This remains true when the sample is split into photometric AGB stars (DLZ$\sim$0.8) and carbon stars, though the evidence for the latter is “barely worth mentioning” (DLZ$\sim$0.2). In contrast, the old population substantially prefers the “rotating” model (DLZ$\sim-0.6$). This may be suggestive of the effects of tidal interactions affecting the old and intermediate-age halo populations differently: we discuss this further below in \S\ref{sec:interactimp}.

\begin{figure*}
\includegraphics[width=0.5\textwidth]{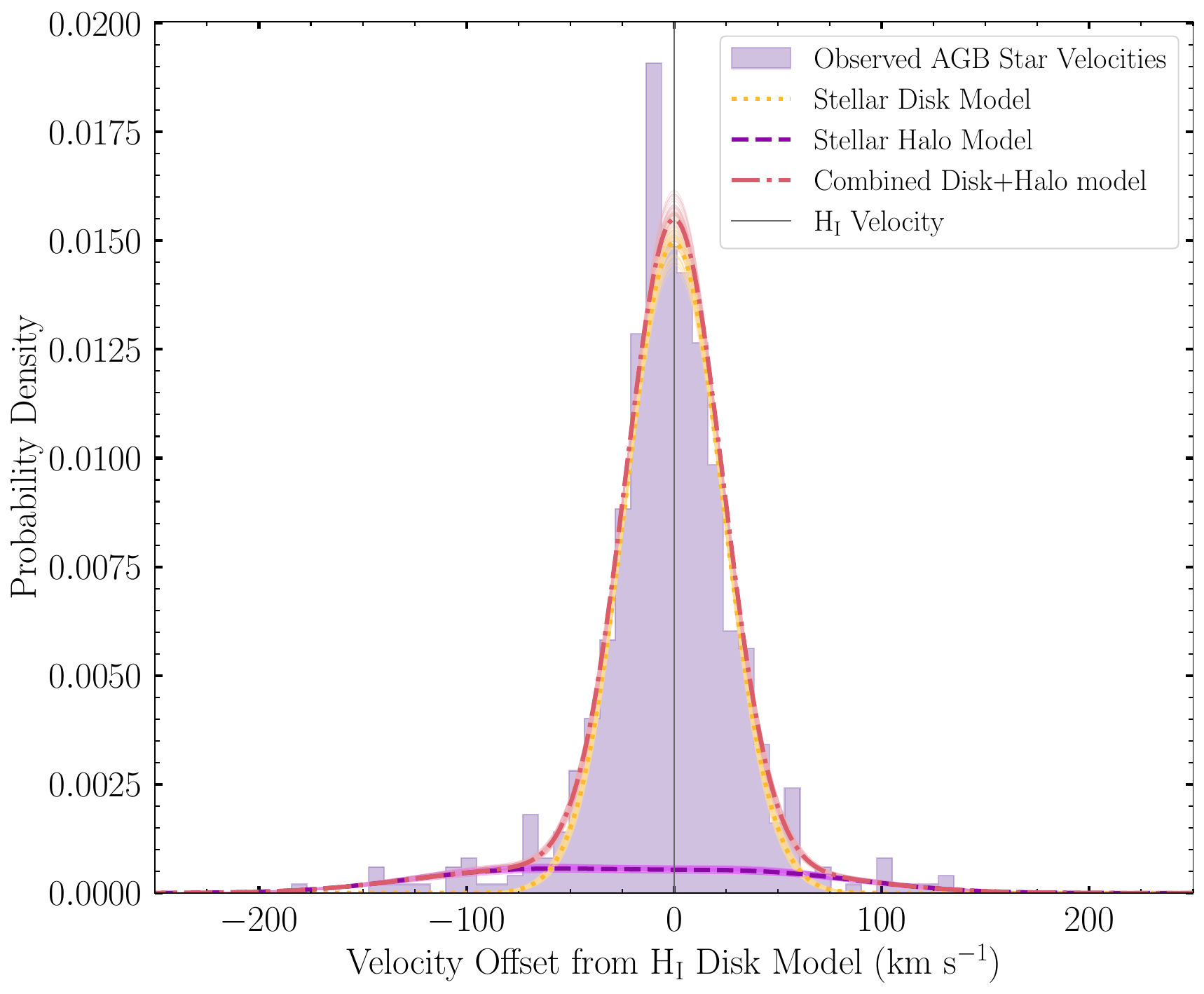}\includegraphics[width=0.5\textwidth]{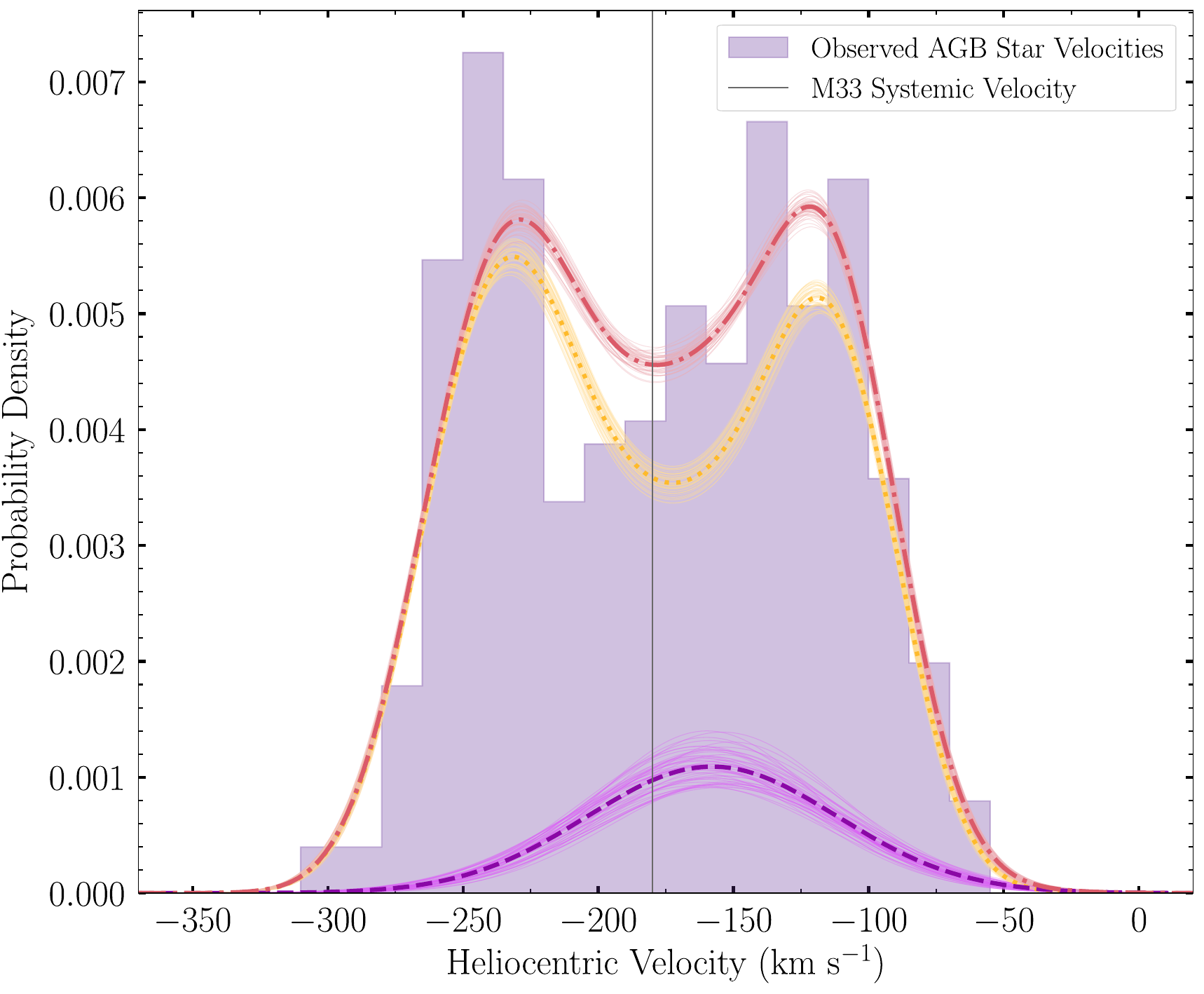}
\caption{As per Fig.~\ref{fig:rotres}, but showing the best-fitting results for the “offset” two-component model -- which allows for a systematic offset from M33's systemic velocity in the higher-dispersion component -- fit to the intermediate-age population. While the best-fit model parameters are consistent between the rotating and offset models, the offset model (for which $\mu_{\text{halo}}=-155$~km s$^{-1}$) is substantially preferred over the rotating model for the intermediate-age population.} 
\label{fig:offres}
\end{figure*}

Fig.~\ref{fig:offres} presents the best-fitting two-component “offset” model for the intermediate-age population, with the associated model parameters given in Table~\ref{tab:agboff}. We find that while the best-fitting values of the model parameters common between the “offset” and “rotating” models (i.e.\ $f_{\text{rot,disk}}, \sigma_{\text{disk}}, f_{\text{halo}}$, and $\sigma_{\text{halo}}$) are all consistent within uncertainty, the systemic offset of the halo-component is inconsistent with zero at the $3\sigma$ level. The preferred systemic velocity of the halo component is $\sim-$155~km~s$^{-1}$; indicating the halo is moving away from us relative to M33’s disk. 

\begin{deluxetable*}{@{\extracolsep{4pt}}cccccccc} \label{tab:agboff}
\tablecaption{Best-fitting parameters for the “offset” two-component model of the intermediate-age population, which allows a systematic offset for the higher-dispersion “halo” population.}
\tablehead{\colhead{}&\colhead{} & \colhead{} & \multicolumn{2}{c}{Disk parameters} & \multicolumn{3}{c}{Halo parameters} \\ \cline{4-5} \cline{6-8}
	\colhead{Radial range} & \colhead{Spatial range} & \colhead{$N_{\text{stars}}$} & \colhead{$f_{\text{rot, disk}}$} & \colhead{$\sigma_{\text{disk}}$} & \colhead{$f_{\text{halo}}$} & \colhead{$\sigma_{\text{halo}}$} & \colhead{$\mu_{\text{halo}}$} }

\startdata
& All & 671 & $0.90^{+0.02}_{-0.02}$ & $22.5^{+0.9}_{-0.9}$ & $0.13^{+0.02}_{-0.02}$ & $49.8^{+5.7}_{-4.8}$ & $-155.9
^{+6.9}_{-7.5}$ \\
All & NE & 338 & $0.89^{+0.02}_{-0.02}$ & $20.2^{+1.8}_{-1.8}$ & $0.15^{+0.05}_{-0.05}$ & $58.0^{+8.4}_{-12.1}$ & $-172.1
^{+25.4}_{-16.1}$ \\
& SW & 333 & $0.90^{+0.03}_{-0.03}$ & $23.4^{+1.5}_{-1.4}$ & $0.16^{+0.06}_{-0.05}$ & $47.4^{+7.4}_{-5.5}$ & $-152.9
^{+9.1}_{-11.3}$ \\ \hline
$<15$ & All & 391 & $0.98^{+0.03}_{-0.04}$ & $24.1^{+1.4}_{-1.3}$ & $0.17^{+0.04}_{-0.03}$ & $41.2^{+5.5}_{-4.6}$ & $-165.8
^{+7.1}_{-7.0}$ \\
$15-30$ & All & 193 & $0.89^{+0.02}_{-0.02}$ & $21.1^{+1.4}_{-1.3}$ & $0.11^{+0.04}_{-0.04}$ & $48.9^{+11.4}_{-8.5}$ & $-158.3
^{+13.7}_{-15.6}$ \\
$30+$ & All & 87 & $0.84^{+0.02}_{-0.02}$ & $15.1^{+1.4}_{-1.3}$ & $0.11^{+0.04}_{-0.04}$ & $62.6^{+18.8}_{-15.6}$ & $-118.2	^{+22.2}_{-24.2}$
\enddata
\tablecomments{Columns are as per Table~\ref{tab:agbrot}, with the exception of $\mu_{\text{halo}}$: this is the central LOS velocity of the halo component in km~s$^{-1}$.}
\end{deluxetable*}

Intriguingly, this is similar to a group of RGB stars in the southern outskirts of M33’s disk (R$\sim$40’), identified by \citet{mcconnachieStellarHaloOuter2006}, which have a mean systemic velocity of $\sim-150$~km~s$^{-1}$. They interpreted this offset group as distinct from M33’s stellar halo, as its dispersion ($\sim$10~km~s$^{-1}$, if poorly constrained) was less than the $\sim$50~km~s$^{-1}$ they expected for M33’s halo based on scaling arguments. Given the relatively small areal coverage and number of stars in their sample, the origin of this population and any potential link to our similarly-offset intermediate-age halo component is unclear. It does, however, raise the question of why such an offset is only seen in our intermediate-age population, and not also preferred for our old RGB population.

\begin{figure}
\includegraphics[height=0.9\textheight]{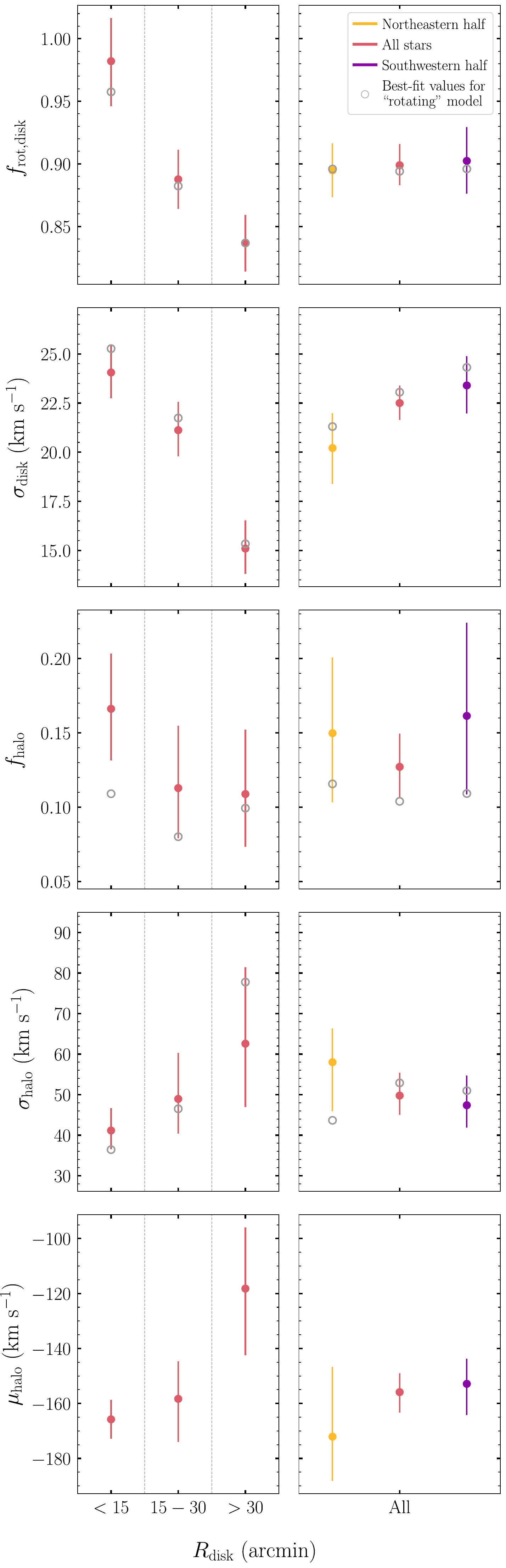}
\caption{As per Fig.~\ref{fig:comprot}, but showing the best-fitting results for the “offset” two-component model fit to the intermediate-age population. Grey circles indicate best-fitting results for the “rotating” two-component model for parameters common to the two models; these are largely consistent within 1$\sigma$, with similar radial trends observed. }
\label{fig:compoff}
\end{figure}

We investigate potential spatial variations associated with the “offset” model by splitting our intermediate-age sample radially and by north/south position as in Section \ref{sec:spatial}. The best-fitting model parameters for each region are included in Table~\ref{tab:agboff}, and Fig~\ref{fig:compoff} compares these parameters to those of the best-fitting “rotating” model for the intermediate-age population in each region. For common parameters between the two models, as for the global fit to the intermediate-age population, these are consistent within uncertainty within each spatial region; accordingly, we also see the same broad spatial trends as observed for the “rotating” model. 

Interestingly, we find that the difference between the mean velocity of the halo component and M33’s systemic velocity (i.e.\ $\mu_{\text{halo}}-v_{\text{sys}}$) is a factor of three larger ($\sim$60~km~s$^{-1}$) in the outermost radial bin than at smaller radii, though this is only inconsistent at the $1\sigma$ level given the large uncertainties on this parameter. As discussed in \S\ref{sec:spatial}, this may also hint at a different perturbed population than the inner halo being fit in this region. However, we do note that due to the limited number of stars in the outermost bin, and the low halo fraction in this region ($\sim$11\%), this result is largely being driven by only $\sim$9 “halo” stars located across the outermost region. We therefore caution over-interpretation of this finding; future analysis of TREX fields located at larger radii is required in order to confirm if this large offset is genuine.

\section{Implications for the origin of M33's halo}\label{sec:implications}
The kinematic structure of M33’s stellar populations allows us insight into the origins of its halo, be this accreted stars from smaller galaxies as expected in $\Lambda$CDM, or stars formed in-situ in M33’s disk and dynamically heated or “kicked up” due to e.g.\ stellar feedback or tidal interactions. In \citetalias{K22}, evidence for both an accreted and in-situ origin is presented. The global lack of rotation for the high-dispersion RGB component may suggest an accreted origin for these stars. However, the close similarity between the photometric metallicity distributions of the disk and halo components, and the decreasing fraction of stars associated with the halo component with radius, may instead point to an in-situ origin. In light of new findings -- including evidence of an extended RGB halo component from \citetalias{smercinaM33STRUCTURE2023}, and our discovery of a substantial ($\sim$10\%) high-dispersion, non-rotating population of intermediate-age stars offset from M33’s systemic velocity, in this section we thus explore potential formation mechanisms for M33’s halo, comparing current expectations for the distinct properties of stellar halos produced via different mechanisms to our observations. While our new measurements unfortunately do not allow us to confidently distinguish between these mechanisms, they do present additional evidence and assist us in determining what additional measurements will be required to do so.

\subsection{Accretion of smaller satellites}\label{sec:accretedimp}
While it is plausible that stars accreted from smaller satellites may contribute to M33's halo, the presence of an intermediate-age halo component makes it very unlikely that this is the only halo formation mechanism in effect.
Halos formed through hierarchical accretion and comprised of debris from low-mass satellites should be both metal-poor \citep{kirbyUniversalStellarMassStellar2013} and old, as these smaller galaxies are likely to be rapidly quenched by e.g.\ reionization \citep{brownQUENCHINGULTRAFAINTDWARF2014} or stripping processes due to interactions with more massive galaxies \citep[e.g.][]{nicholsGasDepletionLocal2011,baheStarFormationQuenching2015}. In the case of massive (i.e. MW-like) galaxies, the accretion history -- and thus debris comprising the stellar halo -- is typically dominated by a single, relatively massive merger event, which sets the average halo metallicity \citep[e.g.][]{dsouzaMassesMetallicitiesStellar2018}. Kinematically, accreted halos are expected to have a high dispersion and display minimal rotation aligned with the disk plane, as debris accreted (approximately) isotropically from many smaller satellites is not expected to mimic the ordered motions of the disk; and while debris from massive mergers has shared integrals of motion linked to the infall kinematics of the progenitor \citep[e.g.][]{ johnstonFossilSignaturesAncient1996,helmiBuildingStellarHalo1999} these are also gradually dispersed over time \citep[e.g.][]{nganDispersalTidalDebris2016, erkalStraySwingScatter2016}. Cosmological models also indicate the fraction of stars associated with an accreted halo component increases significantly with galactocentric radius \citep[e.g.][]{rodriguez-gomezStellarMassAssembly2016,davisonEAGLEViewEx2020}; and simulations of an M33 analogue with an accreted halo by \citet{mostoghiuCLUESM33Reversed2018} suggest this as well. 

Comparing these properties to recent observations of M33, we find some evidence that the old stellar halo component may have an accreted origin. An accreted origin is consistent with the significantly higher dispersion we observe for this component relative to the stellar ($\sim3\times$) and gaseous disks ($\sim7.5\times$: \citealt{cheminAnisotropyRandomMotions2020}) and the lack of (or very minimal) rotation in the plane of the disk (\S\ref{sec:haloprops}). The relative fraction of stars associated with the old halo component we measure matches predictions from \citetalias{smercinaM33STRUCTURE2023} for a power-law profile associated with an accreted halo; and while \citetalias{smercinaM33STRUCTURE2023} note the corresponding global halo fraction is somewhat high to be associated entirely with accreted stars, it is still within the range observed \citep{harmsenDiverseStellarHaloes2017} and simulated \citep[e.g.][]{stewartGasrichMergersLCDM2009,pillepichBUILDINGLATETYPESPIRAL2015, sandersonReconcilingObservedSimulated2018} for more massive MW-like galaxies. However, whether such high fractions are similarly expected for lower-mass galaxies like M33 is not clear, particularly given simulations of significantly lower-mass systems (M$_{\text{DM halo}}<10^{10}$~M$_\odot$) suggest the fraction of accreted stars in such galaxies is negligible \citep{fittsNoAssemblyRequired2018}. Further simulations of accreted halos around M33-mass galaxies are required to confirm if similarly high fractions of accreted stars are expected in this mass range.

However, there are much stronger lines of evidence \textit{against} accretion being the dominant formation mechanism for M33’s halo. Most critical is the presence of a substantial intermediate-age halo component throughout M33: this is entirely unexpected under an accretion scenario, as the low-mass galaxies which comprise accreted halos should not contain significant numbers of these comparatively younger populations. Additionally, we observe a decreasing fraction of old RGB stars associated with the halo component stars at increasing radii -- in direct contrast to predictions from simulations (see further discussion in section 5 of \citetalias{K22}) -- while the fraction of intermediate-age stars associated with the halo remains relatively constant at all radii (\S\ref{sec:varfrac}. This implies the mean age of the halo is \textit{decreasing} with galactocentric radius, in opposition to that expected for accreted halos comprised of purely old stars. 

The photometric metallicity of M33’s halo as measured by \citetalias{K22} also does not align with the expected properties of a purely accreted halo. \citetalias{smercinaM33STRUCTURE2023} derive a total stellar halo mass for M33 of $\sim5\times10^8$~M $_\odot$ given their measured RGB density profile (which agrees with our global $f_{\text{halo}}$ for the old RGB population). Assuming this mass is comprised entirely of accreted debris from a single dominant merger, and comparing to accreted mass-metallicity relations \citep{dsouzaMassesMetallicitiesStellar2018, smercinaRelatingDiverseMerger2022}, predicts a mean halo metallicity of [Fe/H]$\sim-1.4$: somewhat lower than the [Fe/H]$_{\text{phot}}\sim-1$ median halo metallicity from \citetalias{K22}. \citetalias{K22} also find negligible differences in the global photometric [Fe/H] distribution of their RGB disk and halo populations; in contrast, an accreted halo is expected to be more metal-poor on average than the galactic disk.

We do note that the halo mass-metallicity relations discussed above are calibrated using more massive galaxies (i.e.\ $M_{*,\text{tot}}\gtrsim M_{*,\text{tot,MW}}$). Further investigation into how these relations scale for less massive systems -- and whether M33 has experienced the requisite single dominant merger event to follow (a scaled version of) the relation -- is necessary. A spatially-resolved star formation history (SFH) for M33 -- akin to that determined by \citet{lazzariniPanchromaticHubbleAndromeda2022}, but extended to older-age populations -- will assist in determining if such an event has occurred. 
In general, while it is plausible that old, metal-poor stars accreted from smaller satellites may comprise at least some of M33's halo -- \citetalias{K22} note that the mean photometric metallicity of the halo decreases and becomes more distinct from the metallicity distribution of the disk at larger radii, indicating some of these stars may be accreted in origin -- it seems very unlikely that this is the only mechanism contributing to its formation. Spectroscopic metallicity estimates for RGB stars across M33 will assist in identifying potentially accreted stars, particularly at large radii; this possibility is currently being investigated (Cullinane et al. in prep). 

\subsection{“In-situ” mechanisms}\label{sec:insituimp}
The presence of an extended intermediate-age halo component in M33 suggests in-situ mechanisms have contributed substantially to its formation, though the spatially diverse properties of the halo indicate no single mechanism is likely to be entirely responsible.
While there are several different in-situ halo formation mechanisms, each of which leaves distinct chemodynamic signatures on the resultant halo population, there are a number of common properties which distinguish halos produced via these mechanisms from purely accreted halos. Regardless of the associated heating mechanism, halos comprised of dynamically heated disk stars will include stars of a variety of ages and metallicities -- not just old metal-poor stars. This results in a halo which is on average younger, more metal rich, and with a lower alpha-to-iron ratio than a purely accreted halo \citep[e.g.][]{tisseraStellarHaloesSimulated2013,cooperFormationSituStellar2015}. Nonetheless, the halo will still be dominated by relatively old (and hence metal-poor) stars as these have experienced the largest number of perturbations necessary to produce significant heating effects \citep{stinsonFeedbackFormationDwarf2009, el-badryBREATHINGFIREHOW2016}. Halos produced in this way are also expected to be more flattened in shape than accreted halos \citep{mccarthyGlobalStructureKinematics2012,kado-fongSituOriginsDwarf2022}, and to display net prograde rotation \citep{mccarthyGlobalStructureKinematics2012,cooperFormationSituStellar2015}; though \citet{ tisseraStellarHaloesSimulated2013} note a clear age-dependence for this trend, with the oldest dynamically-heated halo stars having no significant rotation.

Unlike for an accreted halo, the substantial intermediate-age halo component we observe is naturally explained via in-situ halo formation. Several different in-situ mechanisms are contenders for the formation of this halo component, but in particular, we posit repeated feedback cycles which drive fluctuations in the global gravitational potential and gradually heat stellar orbits as producing the intermediate-age halo. 

\citet{dobbsSimulationsFlocculentSpiral2018} find strong stellar feedback is needed to match the observed M33 disk properties, and simulations suggest stellar feedback is an effective avenue through which stellar halos in low-mass galaxies can form \citep[e.g.][]{stinsonFeedbackFormationDwarf2009,maxwellBUILDINGSTELLARHALO2012,el-badryBREATHINGFIREHOW2016}. \citet{el-badryBREATHINGFIREHOW2016} suggests a timescale of a few Gyr before coherent, non-reversible heating of stellar orbits occurs: this aligns with the $\sim$3~Gyr mean age of our intermediate-age sample. If similar star formation bursts to those in the recent history of M33 are observed in its more ancient SFH\footnote{existing ancient star-formation histories are based on isolated fields, and not sufficiently precise to observe sub-Gyr SF bursts \citep{williamsDETECTIONINSIDEOUTDISK2009,barkerStarFormationHistory2011}.}, this would provide additional evidence for a feedback-driven origin for M33’s intermediate-age halo. 

While “direct” feedback processes (i.e.\ star formation in outflowing gas) can also produce halo-like components \citep[e.g.][]{stinsonFeedbackFormationDwarf2009,el-badryBREATHINGFIREHOW2016}, the lack of a young ($\sim100$~Myr: \citetalias{Q22}) halo component \citepalias{K22} suggests this particular mechanism contributes much less strongly to the formation of M33’s halo. The recent ($<500$~Myr ago) SFH of M33 does display repeated star-formation bursts \citep{lazzariniPanchromaticHubbleAndromeda2022}, and \citet{el-badryBREATHINGFIREHOW2016} find young stars can respond kinematically to star formation bursts and migrate significantly on timescales of $\sim100$~Myr; accordingly, if this particular mechanism was significant, we would expect to see an even younger M33 halo component which is lacking. 

The repeated cycles of stellar feedback which we posit as forming the intermediate-age halo will similarly affect older stellar populations. As these older stars have experienced even more feedback cycles over their lifetime, they should be even more significantly heated; this may explain the mildly larger dispersion ($\sim$1.2$\times$; a $\sim2\sigma$ increase) we observe for our old halo population relative to the intermediate-age halo. Quantitative comparisons to simulations in terms of the average heating rate and, accordingly, the relative dispersions of different age populations are necessary to determine if the dispersion ratio we observe aligns with that expected for this mechanism. 

In addition, if repeated strong cycles of stellar feedback are responsible for the formation of M33’s halo, this may also explain the decreasing contribution of the old stellar halo at large radii (\S\ref{sec:varfrac}). M33 is thought to have experienced inside-out disk formation \citep[e.g.][]{magriniBuildingDiskGalaxy2007, williamsDETECTIONINSIDEOUTDISK2009, williamsPanchromaticHubbleAndromeda2021}. This means there has been less total star formation (and therefore less feedback) in the outskirts; and, more significantly, the shorter timescale over which star formation has occurred in the outskirts corresponds to a lower total number of inflow-outflow cycles which drive halo formation in this region. The halo density profile break observed at small radii ($\sim2$~kpc/8') by \citetalias{smercinaM33STRUCTURE2023} may similarly reflect the longer, and stronger, contribution from stellar feedback to halo production at small radii. 

The lack of a similar decrease in the intermediate-age halo population fraction with radius is not necessarily inconsistent with this picture: as recent star formation is more extended, the strength and number of inflow-outflow cycles driving halo formation for this comparatively younger population is potentially more consistent across a larger radial range. Dynamical heating from scattering interactions with M33’s weak bar and spiral arms (e.g.\ \citealt{williamsPanchromaticHubbleAndromeda2021}; \citetalias{smercinaM33STRUCTURE2023}) may also contribute to an increased halo fraction for older stellar populations in the inner regions, as these stars have had a longer time period over which to interact with these features and be heated.

Finally, the similar global photometric metallicity distributions measured by \citetalias{K22} for the RGB halo and disk populations are consistent with an in-situ origin: in this scenario, the underlying stellar populations (and thus abundance distribution) of the disk and halo are the same.

We do note, however, that in-situ mechanisms cannot explain every observed property of M33’s halo. Simulations suggest that halos formed from dynamically heated disk stars typically display prograde rotation in the same sense as the galaxy disk \citep{mccarthyGlobalStructureKinematics2012, cooperFormationSituStellar2015}: this is inconsistent with the lack of rotation observed for our intermediate-age halo population. In addition, it seems difficult for stellar feedback on its own to account for the apparent bulk velocity offset between the intermediate-age halo and disk populations, as discussed in \S\ref{sec:offset}. A potential difference in the mean photometric metallicity of the old disk and halo at very large radii, as measured by \citetalias{K22}, is also not obviously explained solely through in-situ mechanisms. 

In general, while repeated cycles of stellar feedback seem likely to contribute significantly to the formation of M33’s halo, this is unlikely to be the sole mechanism at play. We cannot currently rule out other in-situ mechanisms -- such as secular evolution and dynamical heating of disk stars during satellite interactions -- as contributing to the formation of the halo. Further comparisons to simulations of M33-like galaxies, which incorporate all of these effects, are necessary to elucidate the relative contributions of these different mechanisms. 

Simulations do predict that halos formed through dynamical heating processes which simultaneously introduce radial migration (i.e.\ secular evolution processes and repeated cycles of stellar feedback) display net outward radial motions \citep{el-badryBREATHINGFIREHOW2016, kado-fongSituOriginsDwarf2022}. Future measurements of 3D kinematics for stars in M33, which are necessary to detect these radial motions, will help confirm if these mechanisms dominate halo formation\footnote{though we caution that these signatures are likely to be difficult to detect given they can be disrupted by subsequent perturbations including e.g. tidal interactions \citep{ stinsonFeedbackFormationDwarf2009}.}.

\subsection{Tidal Interactions}\label{sec:interactimp}
We observe several potential signatures of tidal interaction in the kinematics of M33's halo populations, most obviously the velocity offset preferred between the intermediate-age disk and halo populations.
Indeed, it has long been clear that M33 has undergone tidal interactions in the past. Its extended \ion{H}{1} disk is warped (e.g.\ \citealt{putmanDISRUPTIONFUELINGM332009, corbelliDynamicalSignaturesLCDMhalo2014}; \citetalias{ kamKinematicsMassDistribution2017}), with the kinematic and spatial distribution of the gas forming an S-shape characteristic of tidal interactions; and there is an equally extended, similarly-shaped low-density stellar structure evident in deep PAndAS photometry \citep{mcconnachieRemnantsGalaxyFormation2009,mcconnachiePhotometricPropertiesVast2010}. More recently, \citetalias{smercinaM33STRUCTURE2023} find a number of structural asymmetries in M33’s inner disk which they attribute to tidal interactions. This includes differences in the strength and orientation of M33’s spiral arms, with the southwestern spiral arm $\sim$50\% stronger and offset in phase by $\sim15^\circ$ relative to the northeastern arm. While the source of these tidal perturbations is currently not well constrained, it seems likely that M33’s halo populations would have been similarly impacted as well. In this section, we discuss potential evidence for M33’s halo having been affected by tidal interactions, and speculate on potential origins for those perturbations. 

The distinct kinematics of both the old and intermediate age halos in the outermost radial bin ($>30’$) relative to those further in -- including a substantially higher dispersion, and in the case of the intermediate-age halo, a much larger velocity offset -- may also be suggestive of tidal interactions perturbing these outermost regions. As mentioned in \S\ref{sec:vardisp}, our observations in the outermost bin are aligned with the region where the \ion{H}{1} disk is beginning to warp; it seems likely that the same interaction which produces the warp may similarly affect the kinematics of stars in this region. Nonetheless, without comparable observations at similar radii perpendicular to the location of the warp, this possibility is difficult to characterize further.  

By far the clearest signature of tidal interactions we observe, however, is the $\sim$25~km s$^{-1}$ velocity offset of the intermediate-age halo relative to the disk. A mechanism which can produce this offset is not immediately apparent. Despite this, we can use the fact that we only observe the offset for the intermediate-age halo (mean age $\sim3$~Gyr), and not the older RGB halo (mean age $>4$~Gyr), to place some very broad constraints on it. In particular, the perturbing interaction must be relatively recent (to avoid leaving imprints on the older stellar populations), but not so recent that the youngest stellar populations (ages $\lesssim200$~Myr) are affected due to the lack of observed halo-like component for these stars \citepalias{K22}. There is some indication of a global enhancement in M33’s SFH over the last 2~Gyr \citep{williamsDETECTIONINSIDEOUTDISK2009}, which broadly aligns with the suggested timing of an interaction based on the mean age of our intermediate-age population. \citetalias{smercinaM33STRUCTURE2023} also observe asymmetries in the spatial distribution of AGB stars of a similar age in M33’s inner regions. 

We do note that the mechanism which produces the offset must, however, be distinct from the type of recent tidal interaction between the LMC and MW which has produced the MW’s halo reflex motion. In that scenario, the shift in COM of the dark matter halo (and corresponding relatively rapid shift of the disk COM to match) produces an apparent velocity offset for \textit{all} stars in the distant halo relative to the disk, regardless of age. 

Regardless of what hypothetical interaction mechanism produces the observed velocity offset for the intermediate-age halo, one obvious contender for the source of the perturbations is M31. Yet recent measurements suggest that M33 is in fact on its first infall to M31 \citep{patelOrbitsMassiveSatellite2017, vandermarelFirstGaiaDynamics2019}, or, less-likely, on a very long-period \citep[$\sim6$~Gyr:][]{patelOrbitsMassiveSatellite2017} orbit; and an M31-M33 interaction occurring at that time -- prior to the formation of most of our intermediate-age sample -- would predominantly affect only older stellar populations. 

However, as pointed out by \citetalias{smercinaM33STRUCTURE2023}, the above calculations of M33’s orbit do not include the effect of M31’s most recent (likely major) merger, which has produced Andromeda’s Giant Stellar Stream \citep{dsouzaAndromedaGalaxyMost2018, hammerBillionYearOld2018}. A major merger could in theory significantly affect M31’s gravitational potential, and thus potentially also the orbit of M33. The current $\sim$10:1 LMC/MW merger significantly affects the MW’s gravitational potential \citep{garavito-camargoQuantifyingImpactLarge2021,lilleengenEffectDeformingDark2023} and by extension the orbits of smaller satellites and stellar streams within it \citep[e.g.][]{erkalTotalMassLarge2019,erkalLimitLMCMass2020,shippMeasuringMassLarge2021}; an up to 2:1 M31 major merger occurring in the past 2-3~Gyr \citep{hammerBillionYearOld2018} would likely have an equal, if not stronger, effect on M31 satellites like M33. Further investigation into the properties and orbit of the GSS progenitor, and its potential impact on M31's gravitational potential and hence M33’s orbit, is required in order to assess the feasibility of this scenario.

What does seem unlikely is a direct interaction between M33 and the GSS progenitor prior to its merging with M31 (regardless of if this was a major or minor merger). Recent models suggest the GSS progenitor likely infalls from M31's northwest \citep{kiriharaFormationAndromedaGiant2017,milosevicMetallicityDistributionProgenitor2022}, which does not align with it having experienced a past interaction with M33 (which, in a first-infall scenario, approaches from M31’s south). 

In summary, while it is obvious that M33 has been affected by tidal interactions, the specifics of these perturbations are as yet unknown. In order to elucidate the source of its perturbed features, further detailed modelling of M33’s past orbit is required. More critically, comparisons with simulations of M33-like galaxies undergoing various types of interactions with different perturbers are necessary to identify possible mechanisms which can produce kinematically offset populations of specific ages, like those we observe. 

\subsection{The localised counter-rotating population}\label{sec:innerrot}
Regardless of what mechanism dominates the formation of M33’s halo, the presence of a localized, counter-rotating RGB halo population in the inner northeast (\S\ref{sec:varhrot}), distinct from the global kinematical trends of the halo as a whole, is not easily explained. In this section, we speculate on a handful of potential explanations, although we caution that each requires a specific finely-tuned scenario, none of which currently have clear supporting evidence. 

There are several proposed “in-situ” mechanisms which can form counter-rotating populations, though these generally produce counter-rotating disk-like (i.e. low-dispersion) populations compared to the high-dispersion population we observe. Bar dissolution is a possibility \citep[e.g.][]{evansSeparatrixCrossingEnigma1994}; however as M33 has an intact and (just) stable bar \citepalias{smercinaM33STRUCTURE2023} this scenario seems unlikely. 

Accretion of counter-rotating gas which subsequently forms counter-rotating stars \citep[e.g.][]{thakarFormationMassiveCounterrotating1996,thakarSmoothedParticleHydrodynamics1998} is also a potential formation mechanism. However, this does not explain why the population is localised in one half of the disk; simulations of this accretion typically produce counter-rotating bulges or disks instead. In addition, as discussed by \citet{corsiniCounterRotationDiskGalaxies2014}, in this scenario the counter-rotating stars should be younger than the existing stellar disk. This is because in order to produce a counter-rotating gaseous disk from which the counter-rotating stars form, there must be minimal co-rotating gas (which can continue to form co-rotating stars) remaining in the system; in gas-rich systems, these two gaseous components collide and lose angular momenta, preventing the formation of counter-rotating stars. There does not, however, appear to be any systemic differences in the CMD distribution of stars in M33’s halo and disk components in this localized region which could indicate potential systemic age differences, though the relatively low number of individual stars in this population makes this comparison difficult. A full census of the stars in the counter-rotating population across M33 may provide sufficient numbers for broad SFH derivations, which can be compared to those for M33’s disk \citep[e.g.][]{lazzariniPanchromaticHubbleAndromeda2022} in order to establish if there is a systematic mean age difference for this population. 

Minor \citep[e.g.][]{bassettFormationS0Galaxies2017} or major mergers \citep{puerariFormationMassiveCounterRotating2001,martelFormationCounterrotatingStars2020} under very specific conditions can also produce counter-rotating populations, both from their own accreted stars, and new stars formed in gas accreted from the merger progenitor. However, as above, the distribution of these counter-rotating populations is typically not localized to just one region of the disk. It is, however, nominally possible that a single small satellite could be accreted on an orbit and with such timing so to form a counter-rotating population that currently dominates the halo in a localized region. In any of these scenarios, the metallicity and $\alpha$-abundance of the counter-rotating stars should also differ from the dominant disk population; spectroscopic abundance measurements will be necessary to determine if this is the case.

\section{Summary} \label{sec:concs}
Using the TREX survey, we have studied the stellar kinematics of old (RGB) and intermediate-age (photometrically selected AGB and spectroscopically identified carbon stars) populations across M33. We model the stellar LOS kinematics relative to those of M33’s gaseous \ion{H}{1} disk (\S\ref{sec:maths}), finding decisive evidence for two distinct components in populations of both ages: a low-dispersion ($\sim22$~km s$^{-1}$) disk-like component co-rotating with the gas, and a significantly higher-dispersion component ($\sim50-60$~km~s$^{-1}$) which does not strongly rotate in the same plane as the gas. We nominate this higher-dispersion component as M33's stellar halo. The halo is dominated by older stars, but there remains a non-trivial contribution from intermediate-age stars -- globally, $\sim$25\% of our old stellar sample and $\sim$10\% of our intermediate-age sample are associated with the halo. This is the first such investigation into the kinematics of AGB stars, and expands on analysis of RGB stars in M33 by \citetalias{K22} by using a $\sim$22\% larger and more uniformly spatially distributed sample. 

We subsequently investigate potential spatial variations in the properties of the two components. We find that within each (radial and north/south-split) region, the best-fitting parameters for the old and intermediate-age populations (excepting $f_{\text{halo}}$) are consistent within uncertainty. There is a mild decrease in both the disk dispersion and fractional rotation speed of the disk relative to that of the \ion{H}{1} gas with increasing radius, in general agreement with recent results from \citetalias{Q22}. In the halo, we find that for a given age population, the best-fitting dispersion is generally consistent within radii $<30$’, but is significantly increased for stars at radii $>30$’, which includes stars beyond the break in M33’s surface brightness profile. Given the spatial distribution of our sample at these radii is dominated by stars in the vicinity of M33’s gaseous and stellar warp, we suggest this population may instead describe debris perturbed by tidal interactions, and not a smooth continuation of the inner halo. Increased spectroscopic coverage of the region beyond 30’, at all position angles, is required to constrain the origin of these stars. 

The fraction of each population associated with the halo component displays different spatial trends for the two age groups: this is consistent across each spatial sub-region for the intermediate-age population, but decreases significantly with radius and is $\sim2\sigma$ lower in M33's northeast for the old population. The decreasing trend with radius for the old population aligns with kinematic models from \citetalias{K22} and fits to the radial density profile of RGB stars from \citetalias{smercinaM33STRUCTURE2023}. 

We also find a localized high-dispersion population of old RGB stars in M33’s inner northeastern disk ($R<15$’) which appear to be strongly counter-rotating relative to the gaseous and stellar disk in this region, confirming similar results from \citetalias{K22}. Additionally, we find indications there may be an intermediate-age population with similar kinematics in the same region. Further investigation is necessary to better characterize the properties of this population and its origin; most potential in-situ formation mechanisms predict such counter-rotating populations should have differing ages or abundances than those of the nearby disk.

We find positive evidence that the intermediate-age halo population is systematically offset from the systemic velocity of M33 by $\sim25$~km~s$^{-1}$, with a preferred central LOS velocity of $\sim-155$~km~s$^{-1}$: indicating this population is moving towards us relative to M33. The origin of this offset is unclear, particularly given the lack of evidence for a similar offset in the old stellar halo velocity, but we posit a relatively recent ($2-3$~Gyr ago) tidal interaction could nominally be responsible. Further modelling of M33's orbit, and comparison to simulated analogues, is necessary to determine if this is a plausible scenario, and the source of the perturbation if so. 

A variety of mechanisms likely influence the formation and evolution of M33’s halo populations. 
In-situ mechanisms which act to dynamically heat disk stars into a halo configuration can naturally explain the presence of an intermediate-age halo; in contrast, accreted halos are dominated by the old, metal-poor stars which comprise the smaller systems from which they originate. Conversely, a global lack of co-rotation with the \ion{H}{1} disk is suggestive of an accreted origin; simulations predict disk stars heated into a halo configuration via in-situ mechanisms should retain prograde rotation. A decreasing halo fraction with radius for the old halo population does not match predictions for accreted halos in cosmological simulations, but this trend is potentially explainable if in-situ mechanisms dominate halo formation in the inner regions of M33. If repeated cycles of stellar feedback act to gradually dynamically heat disk stars, this effect will be strongest for the old stars in the inner galaxy where star formation is concentrated, contributing to a greater fraction of stars associated with the old halo at small radii. In-situ mechanisms also explain the similar photometric metallicity distributions for the old disk and halo populations in the inner regions of M33 measured by \citetalias{K22}. Improved modelling of M33’s stellar kinematics independent to that of the gas, better characterisation of the spatial distribution of different kinematical components, and spectroscopically-derived abundance measurements will each help constrain the origin of these different populations, as well as provide further insight into the processes which shape disk evolution and stellar halo formation in the regime of high-mass dwarf galaxies. 


\begin{acknowledgments}
We thank Chiara Villanueva and Taylor Gorkos for their contributions to analysis of factors potentially impacting the measured disk dispersion. The authors recognize and acknowledge the very significant cultural role and reverence that the summit of Mauna Kea has always had within the indigenous Hawaiian community. We are most fortunate to have the opportunity to conduct observations from this mountain. Support for this work was provided by NSF grant AST-1909066 (K.M.G. and L.R.C.). I.E. acknowledges generous support from a Carnegie-Princeton Fellowship through Princeton University. The analysis pipeline used to reduce the DEIMOS data was developed at UC Berkeley with support from NSF grant AST-0071048.
\end{acknowledgments}

\facilities{Keck:II(DEIMOS), CFHT(MegaCam), HST(ACS)}
\software{astropy \citep{astropycollaborationAstropyCommunityPython2013,astropycollaborationAstropyProjectBuilding2018a,astropycollaborationAstropyProjectSustaining2022}, corner \citep{foreman-mackeyCornerPyScatterplot2016}, dustmaps \citep{greenDustmapsPythonInterface2018}, dynesty v2.0.1 \citep{speagleDynestyDynamicNested2020,koposovJoshspeagleDynestyV22022}, matplotlib \citep{hunterMatplotlib2DGraphics2007}, numpy \citep{vanderwaltNumPyArrayStructure2011}} 
\bibliographystyle{aasjournal}
\bibliography{AGBhalo.bib,software.bib}


\end{document}